\documentclass[preprint,prd,nofootinbib,tightenlines,amsmath]{revtex4}
\usepackage{enumerate}
\usepackage{multirow}
\usepackage[utf8]{inputenc}
\usepackage{anysize}
\usepackage{textcomp}
\usepackage{epsfig}
\usepackage{graphics,color} 
\usepackage{verbatim}
\usepackage{afterpage}
\usepackage{amsmath}    
\usepackage{soul}
\usepackage{xcolor,cancel}

\DeclareUnicodeCharacter{00A0}{ }

\usepackage{blindtext}

\catcode`\@=11
\def\sla@#1#2#3#4#5{{%
 \setbox\z@\hbox{$\m@th#4#5$}%
 \setbox\tw@\hbox{$\m@th#4#1$}%
 \dimen4\wd\ifdim\wd\z@<\wd\tw@\tw@\else\z@\fi
 \dimen@\ht\tw@
 \advance\dimen@-\dp\tw@ \advance\dimen@-\ht\z@
 \advance\dimen@\dp\z@
 \divide\dimen@\tw@ \advance\dimen@-#3\ht\tw@
 \advance\dimen@-#3\dp\tw@ \dimen@ii#2\wd\z@
 \raise-\dimen@\hbox to\dimen4{%
 \hss\kern\dimen@ii\box\tw@\kern-\dimen@ii\hss}%
 \llap{\hbox to\dimen4{\hss\box\z@\hss}}}}

\def\cpto{\mathrel {\vcenter {\baselineskip 0pt \kern 0pt
    \hbox{$H_{r.f.}$} \kern 0pt \hbox{$\longrightarrow$} }}}
\def\slashed#1{%
 \expandafter\ifx\csname sla@\string#1\endcsname\relax
{\mathpalette{\sla@/00}{#1}}
\fi}
\def\declareslashed#1#2#3#4#5{%
 \expandafter\def\csname sla@\string#5\endcsname{%
#1{\mathpalette{\sla@{#2}{#3}{#4}}{#5}}}}
 \catcode`\@=12
\declareslashed{}{/}{.08}{0}{D}
 \declareslashed{}{/}{.1}{0}{A}
 \declareslashed{}{/}{0}{-.05}{k}
 \declareslashed{}{/}{.1}{0}{\partial}
 \declareslashed{}{\not}{-.6}{0}{f}

\def\lsim{\mathrel {\vcenter {\baselineskip 0pt \kern 0pt
    \hbox{$<$} \kern 0pt \hbox{$\sim$} }}}
\def\gsim{\mathrel {\vcenter {\baselineskip 0pt \kern 0pt
    \hbox{$>$} \kern 0pt \hbox{$\sim$} }}}

\newcommand{\bea}{\begin{eqnarray}}
\newcommand{\eea}{\end{eqnarray}}

\begin{document}

\baselineskip=15pt
\preprint{}

\title{Validity of two Higgs doublet models with a scalar color octet up to a high energy scale}

\author{Li Cheng$^{1}$\footnote{Electronic address: lcheng@iastate.edu} and German Valencia$^{2}$\footnote{Electronic address: German.Valencia@monash.edu }}


\affiliation{$^{1}$ Department of Physics and Astronomy, Iowa State University, Ames, IA 50011.}

\affiliation{$^{2}$ School of Physics and Astronomy, Monash University, Melbourne, Australia.}
\date{\today}

\vskip 1cm
\begin{abstract}

We have recently studied theoretical constraints on the parameters of a 2HDM augmented with a color-octet scalar. In this paper we consider the consequences of requiring the model to remain valid up to very high energy scales, such as the GUT scale. The acceptable region of parameter space is reduced when one insists on vacuum stability, perturbative unitarity and the absence of Landau poles below a given scale. As the scale to which we require the model to be valid is increased, the acceptable region of parameter space for the 2HDM sector is reduced in such a way that it approaches the alignment limit, $\cos(\beta-\alpha)\to 0$, and the masses of $H^0$, $A$ and $H^\pm$ are pushed closer to each other. The parameters of the color octet sector are also restricted to an increasingly smaller region.

\end{abstract}

\maketitle

\section{Introduction}

An interesting extension of the standard model (SM) proposed by Manohar and Wise \cite{Manohar:2006ga} consists of adding a color-octet electroweak doublet scalar. The original motivation for this extension is that it consists of one of two scalar representations  allowed by minimal flavor violation for scalars that do not transform under the flavor group. From this perspective the most general renormalizable potential consistent with this field content is constructed resulting in generic studies for TeV scale color octet scalars. Scalars such as these occur in specific models, for example in unification with $SU(5)$ \cite{Perez:2016qbo, Dorsner:2007fy} or with $SO(10)$ \cite{Bertolini:2013vta}.  
The model has interesting phenomenology and many studies have been performed in the literature \cite{Gresham:2007ri,Gerbush:2007fe,Burgess:2009wm,He:2011ti,Dobrescu:2011aa,Bai:2011aa,Arnold:2011ra,Kribs:2012kz,Reece:2012gi,Cao:2013wqa,He:2013tla,Cheng:2015lsa,Martinez:2016fyd}. A salient feature of the MW model being that it is very difficult to observe or exclude the new scalars at the LHC. The model has a very large parameter space, but these studies have shown that it is possible to constrain it significantly by imposing custodial symmetry, partial wave unitarity and vacuum stability.

Along these lines, we have recently proposed considering the MW extension in the context of a two Higgs doublet model (2HDM). In Ref.~\cite{Cheng:2016tlc} we introduced the model and constrained its parameter space by imposing tree level theoretical constraints arising from symmetries and from perturbative unitarity. In addition we briefly discussed LHC phenomenology, concluding that the largest effects from this addition to the 2HDM would appear in corrections to one-loop couplings of the higgs boson to two gluons or photons. 

Previous studies of the MW model \cite{He:2013tla} and of the 2HDM \cite{Chakrabarty:2014aya,Ferreira:2015rha} indicate that the viable parameter space is further constrained when one includes renormalization group corrections, and in this paper we apply this rationale to the MW extension of the 2HDM. We first compute the beta functions for the couplings of the model.  We then consider the requirements that there be no Landau poles (LP) below a certain high mass scale $\Lambda$,  that the scalar potential remains stable and that two-to-two scattering amplitudes remain perturbative at all scales below $\Lambda$. These conditions have a long history of being used as theoretical constraints from their early application to the SM Higgs boson mass \cite{Callaway:1988ya,Sher:1988mj}.

\section{The model}

The construction of the model was described in Ref.~\cite{Cheng:2016tlc}. The scalar sector of the SM is replaced by three $SU(2)$ scalar doublets: two color singlets ($\Phi_1, \Phi_2)$ and one color-octet $S$. The most general renormalizable potential for the 2HDM sector ($\Phi_1, \Phi_2)$ is well known from the literature \cite{Gunion:1989we,Branco:2011iw} and we restrict ourselves to the case of a CP conserving potential with a discrete symmetry $\Phi_1\to -\Phi_1$ that is only violated softly (although in the end we replace this with the requirement of MFV). To this known potential we add the couplings between the color octet $S$ and the two color singlets ($\Phi_1, \Phi_2)$ as well as the color octet self interactions \cite{Manohar:2006ga} resulting in
\begin{eqnarray}
V\left( \Phi_1, \Phi_2 \right)  &=& m_{11}^2 \Phi_1^\dag \Phi_1 + m_{22}^2 \Phi_2^\dag \Phi_2 - m_{12}^2 \left( \Phi_1^\dag \Phi_2 + \Phi_2^\dag \Phi_1 \right) \nonumber \\
&+& \frac{\lambda_1}{2} \left( \Phi_1^\dag \Phi_1 \right)^2 + \frac{\lambda_2}{2} \left( \Phi_2^\dag \Phi_2 \right)^2 
+ \lambda_3 \left( \Phi_1^\dag \Phi_1 \right) \left( \Phi_2^\dag \Phi_2 \right)  \nonumber\\
&+& \lambda_4 \left( \Phi_1^\dag \Phi_2 \right) \left( \Phi_2^\dag \Phi_1 \right) + \frac{\lambda_5}{2} \left[ \left( \Phi_1^\dag \Phi_2 \right)^2 + \left( \Phi_2^\dag \Phi_1 \right)^2 \right] \nonumber\\
&+& 2m_S^2 {\rm Tr}S^{\dag i}S_i + \mu_1 {\rm Tr}S^{\dag i}S_i S^{\dag j}S_j + \mu_2 {\rm Tr}
S^{\dag i}S_j S^{\dag j}S_i + \mu_3 {\rm Tr} S^{\dag i}S_i {\rm Tr}S^{\dag j} S_j\nonumber\\
& +& \mu_4 {\rm Tr}S^{\dag i}S_j {\rm Tr}S^{\dag j}S_i + \mu_5 {\rm Tr}S_i S_j{\rm Tr}
S^{\dag i}S^{\dag j} + \mu_6 {\rm Tr}S_i S_j S^{\dag j}S^{\dag i}\nonumber \\
&+& \nu_1 \Phi_1^{\dag i}\Phi_{1i}{\rm Tr}S^{\dag j}S_j + \nu_2 \Phi_1^{\dag i}\Phi_{1j}
{\rm Tr}S^{\dag j}S_i\nonumber\\
& +& \left( \nu_3 \Phi_1^{\dag i}\Phi_1^{\dag j}{\rm Tr}S_i S_j + \nu_4 \Phi_1^{\dag i}{\rm Tr}
S^{\dag j}S_j S_i + \nu_5 \Phi_1^{\dag i}{\rm Tr}S^{\dag j}S_i S_j + {\rm h.c.} \right) \nonumber \\
&+&\omega_1 \Phi_2^{\dag i}\Phi_{2i}{\rm Tr}S^{\dag j}S_j + \omega_2 \Phi_2^{\dag i}\Phi_{2j}
{\rm Tr}S^{\dag j}S_i\nonumber\\
& +& \left( \omega_3 \Phi_2^{\dag i}\Phi_2^{\dag j}{\rm Tr}S_i S_j + \omega_4 \Phi_2^{\dag i}{\rm Tr}
S^{\dag j}S_j S_i + \omega_5 \Phi_2^{\dag i}{\rm Tr}S^{\dag j}S_i S_j + {\rm h.c.} \right) \nonumber\\
&+& \kappa_1 \Phi_1^{\dag i}\Phi_{2i}{\rm Tr}S^{\dag j}S_j +
\kappa_2 \Phi_1^{\dag i}\Phi_{2j}{\rm Tr}S^{\dag j}S_i + \kappa_3 \Phi_1^{\dag i}\Phi_2^{\dag j}{\rm Tr}S_j S_i
+ {\rm h.c.}
\label{potential}
\end{eqnarray}
In all the terms we have explicitly shown the  $SU(2)$ indices $i, j$, $S_i = T^A S_i^A$, and the trace is taken over color indices.
As per our discussion in Ref.~\cite{Cheng:2016tlc}, in terms that are not part of the usual the 2HDM, we have allowed some that satisfy MFV but not  the discrete symmetry mentioned above. 
After symmetry breaking some of these couplings are related to scalar masses, and these relations can be readily found in the literature \cite{Manohar:2006ga,Branco:2011iw,Cheng:2016tlc}.

The Yukawa potential for this model  in the flavour eigenstate basis is given by
\begin{eqnarray}
L_{Y}&=&  - {\left( g_1^D \right)^\alpha}_\beta {\bar D}_{R, \alpha }\Phi_1^\dag Q_L^\beta -
{\left( g_1^U \right)^\alpha}_\beta {\bar U}_{R, \alpha}{\tilde \Phi}_1^\dag Q_L^\beta \nonumber\\
&-& {\left( g_2^D \right)^\alpha}_\beta {\bar D}_{R, \alpha } \Phi_2^\dag Q_L^\beta  - {\left( g_2^U
\right)^\alpha}_\beta {\bar U}_{R, \alpha}{\tilde \Phi}_2^\dag Q_L^\beta + {\rm h.c.},\nonumber \\
&-& {\left( g_3^D \right)^\alpha}_\beta {\bar D}_{R, \alpha}S^\dag Q_L^\beta  -
{\left( g_3^U \right)^\alpha}_\beta {\bar U}_{R, \alpha }{\tilde S}^\dag Q_L^\beta  + {\rm h.c.}
\label{yukawas}
\end{eqnarray}
As is conventional, we use ${\tilde H}_i = \varepsilon_{ij} H_j^*$
 for all three scalar doublets $H=\Phi_{1,2},S$,  and $\alpha, \beta$ are flavour indices.

The large number of parameters present in Eq.~\ref{potential} and Eq.~\ref{yukawas} is reduced by the following theoretical considerations:
\begin{itemize}
\item Minimal flavour violation \cite{Chivukula:1987py,D'Ambrosio:2002ex} implies
\begin{itemize}
\item  2HDM Type I: $\eta_1^D=\eta_1^U =0$ 
\item 2HDM Type II: $\eta_1^U=\eta_2^D=0$
\end{itemize}
Requiring MFV instead of a discrete symmetry to define the models allows quartic terms in the scalar potential that are odd in either of the doublets, such as $\nu_{4,5}$, $\omega_{4,5}$ and $\kappa_{1,2,3}$. It also allows additional terms in the pure 2HDM sector, but we do not consider those here. 

\item Custodial symmetry \cite{Sikivie:1980hm,Pomarol:1993mu,Grzadkowski:2010dj}. As discussed in \cite{Cheng:2016tlc} the least restrictive method to impose custodial symmetry results in all the  $\lambda_i$'s being real and in the relations
\begin{eqnarray}
{\kappa_2} = {\kappa _3},\ 2{\nu _3} = {\nu _2},\ {\nu _4} = \nu _5^*,\ 2{\omega _3} = {\omega_2},\ {\omega _4} = \omega _5^*,\ \lambda_4 = \lambda_5.
\label{met1}
\end{eqnarray}
These conditions imply mass degeneracies $m_{H^\pm}=m_A$ and $m_{S^\pm}=m_{S_I^0}$. An alternative possibility,  `twisted' custodial symmetry \cite{Gerard:2007kn,Cervero:2012cx} results instead in  $m_{H^\pm}=m_H$ and $m_{S^\pm}=m_{S_R^0}$ \cite{Burgess:2009wm}.

\item Perturbative unitarity \cite{Lee:1977eg,Kanemura:1993hm, Horejsi:2005da,Ginzburg:2005dt,Grinstein:2015rtl,He:2013tla} 

The two-to-two scattering matrix for this model in the neutral, color singlet channel is $18\times 18$ and in Ref.~\cite{Cheng:2016tlc} we diagonalize it numerically and illustrate the resulting constraints. The numerical results can be roughly approximated by,
\begin{subequations}
\begin{align}
&\left| \lambda_1 \right|,\left| \lambda_2 \right| \le \frac{8\pi}{3},\quad
\left| \lambda_3 \right| \le 4\pi,\quad \left| \lambda_4 \right|, \left| \lambda_5 \right| \le \frac{8\pi}{5},\\
&\left| \nu_1 \right|,\left| \nu_3 \right|, \left| \omega_1 \right|, \left| \omega_3 \right| \le 2\sqrt{2}\pi,\quad
\left| \nu_2 \right|,\left| \omega_2 \right| \le 4\sqrt{2}\pi,\\
&\left| \kappa_1 \right| \le 2\pi,\quad \left| \kappa_2 \right|, \left| \kappa_3 \right| \le 4\pi,\\
& \left| 17 \mu_3 +13 \mu_4 +13 \mu_6 \right|\leq 16 \pi,\\
& \left| \nu_4+\nu_5\right|\leq \frac{32\pi}{\sqrt{15}}, \quad 
 \left| \omega_4+\omega_5\right| \leq  \frac{32\pi}{\sqrt{15}},\\
& \left| 12 \mu_3+10 \mu_4+7\mu_6\right|\leq 32\pi.
\label{approxlim}
\end{align}
\end{subequations}
In the present study we require the perturbative unitarity constraints, as obtained by diagonalizing the full $18\times 18$ matrix numerically, to be satisfied by the running couplings at all scales below $\Lambda$.

\item Stability, or having a positive definite Higgs potential for the 2HDM  \cite{Deshpande:1977rw} implies that 
\begin{equation}
\lambda_1 > 0, \quad
\lambda_2 > 0, \quad
\lambda_3 > - \sqrt{\lambda_1 \lambda_2},\quad
\lambda_3 + \lambda_4 \pm \lambda_5 > - \sqrt{\lambda_1 \lambda_2}.
\label{stabvac}
\end{equation}
We will again require that this conditions be satisfied by the running couplings at all scales below $\Lambda$. A recent paper presents a way to extend the stability conditions to additional fields \cite{Kannike:2016fmd}. Here we will only consider the effect of the color scalars through their one-loop contributions to the running of the parameters of the 2HDM, and require the conditions Eq.~\ref{stabvac} be satisfied at all scales.

\item For our numerical analysis we will use color octet scalar masses near 1~TeV as in Ref.~\cite{Cheng:2016tlc}, as masses at this scale are still allowed by LHC searches \cite{Hayreter:2017wra}.

\end{itemize}

\section{Validity of the model up to high energy scales} 

\subsection{The renormalization group equations for the scalar couplings}

We now turn to the novel aspect of this paper: investigating the consequences of requiring the model to be valid up to some high energy scale. The procedure is straightforward, we first derive the corresponding renormalization group equations (RGE) for all the parameters in the model, a result we present in the Appendix. For our numerical analysis we restrict ourselves to the case with custodial symmetry and no CP violation, which reduces the RGE to the following,

\everymath{\displaystyle}

\begingroup
\allowdisplaybreaks
\begin{align*}
&16\pi^2\beta_{\lambda_1} = 12\lambda_1^2 + 4\lambda_3^2 + 4\lambda_3\lambda_4 +4\lambda_4^2 
+ 8\nu_1^2 + 8\nu_1\nu_2 + 8\nu_2^2 \\
&\qquad\qquad\quad - 12 \lambda_t^4 - 3\lambda_1\left( 3g^2 - 4\lambda_t^2 + g'^2 \right) + \frac34 \left( 3g^4 + 2g^2 g'^2 + g'^4 \right),\\
&16\pi^2\beta_{\lambda_2} = 12\lambda_2^2 + 4\lambda_3^2 + 4\lambda_3\lambda_4 + 4\lambda_4^2 
+ 8\omega_1^2 + 8\omega_1\nu_2 + 4\omega_2^2 + 16\omega_3^2,\\
&\qquad\qquad\quad - 3\lambda_2\left( 3g^2 + g'^2 \right) + \frac34 \left( 3g^4 + 2g^2 g'^2 + g'^4 \right),\\
&16\pi^2\beta_{\lambda_3} = 4\lambda_3^2 + 4\lambda_4^2 + 2\left(\lambda_1+\lambda_2\right)\left(3\lambda_3 + \lambda_4\right)
+ 8\nu_1\omega_1 + 4\nu_1\omega_2 + 4\nu_2\omega_1 + 8\kappa_2^2,\\
&\qquad\qquad\quad  - 3\lambda_3\left( 3g^2 - 2\lambda_t^2 + g'^2 \right) + \frac34 \left( 3g^4 - 2g^2 g'^2 + g'^4 \right),\\
&16\pi^2\beta_{\lambda_4} = 12\lambda_4^2 + 8\lambda_3\lambda_4 + 2\left(\lambda_1+\lambda_2\right)\lambda_4 + 4\nu_2\omega_2
+ 8\left|\kappa_1\right|^2 + 8\kappa_1\kappa_2 + 4\kappa_2^2,\\
&\qquad\qquad\quad - 3\lambda_4\left( 3g^2 - 2\lambda_t^2 + g'^2 \right) + \frac32 g^2 g'^2,\\
&16\pi^2\beta_{\nu_1} = 6\lambda_1\nu_1 + 2\lambda_1\nu_2 + 4\lambda_3\omega_1 + 2\lambda_3\omega_2 + 2\lambda_4\omega_1 + 2\nu_1^2 + 2\nu_2^2 + 2\kappa_1^2 + 2\kappa_2^2 \\
&\qquad\qquad\quad + \nu_1\left(26\mu_1 + 17\mu_3 + 13\mu_4\right)
+\nu_2\left(\frac{32}3\mu_1 + 8\mu_3 + 2\mu_4\right) + \frac{10}3 \nu_4^2,\\
&16\pi^2\beta_{\nu_2} = 2\lambda_1\nu_2 + 2\lambda_4\omega_2 + 4\nu_1\nu_2 + 6\nu_2^2  + 4\kappa_1\kappa_2 + 6\kappa_2^2 \\
&\qquad\qquad\quad + \nu_2\left(\frac{14}3\mu_1 + \mu_3 + 9\mu_4\right) + \frac{25}3\nu_4^2,\\
&16\pi^2\beta_{\nu_4} =\left(3\kappa_1 + 9\kappa_2\right)\omega_4 + \nu_4\left(3\nu_1 + 9\nu_2 + 11\mu_1 + 3\mu_3 + 9\mu_4\right),\\
&16\pi^2\beta_{\omega_1} = 6\lambda_2\omega_1 + 2\lambda_2\omega_2 + 4\lambda_3\nu_1 + 2\lambda_3\nu_2 + 2\lambda_4\nu_1 + 2\omega_1^2 + 2\omega_2^2 + 2\kappa_1^2 + 2\kappa_2^2 \\
&\qquad\qquad\quad + \omega_1\left(26\mu_1 + 17\mu_3 + 13\mu_4\right)
+\omega_2\left(\frac{32}3\mu_1 + 8\mu_3 + 2\mu_4\right) + \frac{10}3\omega_4^2,\\
&16\pi^2\beta_{\omega_2} = 2\lambda_2\omega_2 + 2\lambda_4\nu_2 + 4\omega_1\omega_2 + 6\omega_2^2 + 4\kappa_1\kappa_2 + 6\kappa_2^2 \\
&\qquad\qquad\quad+ \omega_2\left(\frac{14}3\mu_1 + \mu_3 + 9\mu_4\right) + \frac{25}3\omega_4^2,\\
&16\pi^2\beta_{\omega_4} = \left(3\kappa_1 + 9\kappa_2\right)\nu_4 + \omega_4\left(3\omega_1 + 9\omega_2 + 11\mu_1 + 3\mu_3 + 9\mu_4\right),\\
&16\pi^2\beta_{\kappa_1} = \kappa_1\left(2\lambda_3 + 10\lambda_4 + 2\nu_1 + 2\omega_1 + 18\mu_1  + 17\mu_3 + 13\mu_4 \right)\\
&\qquad\qquad\quad +\kappa_2\left(4\lambda_4 + \frac32\nu_2 + \frac32\omega_2 + \frac{32}3\mu_1
+ 8\mu_3 + 4\mu_4\right) + \frac{10}3\nu_4\omega_4,\\
&16\pi^2\beta_{\kappa_2} =  \kappa_1\left(2\nu_2 + 2\omega_2 \right) + 14\nu_4\omega_4\\
&\qquad\qquad\quad+\kappa_2\left(2\lambda_3 + 2\lambda_4 + 2\nu_1 + 4\nu_2 + 2\omega_1 + 4\omega_2 + \frac{14}3\mu_1 + \mu_3 + 9\mu_4\right) ,\\
&16\pi^2\beta_{\mu_1} =  3\nu_4^2 + 3\omega_4^2
+ 13\mu_1^2 + 6\mu_1\left(\mu_3 + \mu_4\right),\\
&16\pi^2\beta_{\mu_3} = 2\nu_1^2 + 2\nu_1\nu_2 + 2\omega_1^2 + 2\omega_1\omega_2 + 4\kappa_1^2 + 4\kappa_1\kappa_2 - \frac{10}3\left(\nu_4^2 + \omega_4^2\right)\\
&\qquad\qquad\quad + \frac{268}9\mu_1^2  + \mu_1\left(52\mu_3 + \frac{88}3\mu_4\right)
+ 20\mu_3^2 + 26\mu_3\mu_4 + 6\mu_4^2,\\
&16\pi^2\beta_{\mu_4} = \nu_2^2 + \omega_2^2 + 2\kappa_2^2 + \frac23\left(\nu_4^2 + \omega_4^2 \right) + \frac49\mu_1^2 + \frac{52}3\mu_1\mu_4 + 6\mu_3\mu_4 + 16\mu_4^2,\\
&16\pi^2\beta_{g_s} = - 6g_s^3,\\
&16\pi^2\beta_g = - \frac53 g^3,\\
&16\pi^2\beta_{g'} = \frac{25}3 g'^3,\\
&16\pi^2\beta_{\lambda_t} = \lambda_t\left(-\frac94 g^2 - 8g_s^2 + \frac{13}2\lambda_t^2
- \frac{17}{12}g'^2\right).
\end{align*}
\endgroup

In these equations $g$, $g'$ and $g_s$ are the $SU(2)_L$, $U(1)_Y$ and $SU(3)$ couplings of the SM, $\lambda_t \equiv \sqrt2 m_t / v$ is the top-quark Yukawa coupling and we use the standard definition $\beta=\frac{\rm d}{{\rm d}\ln\left(\Lambda/\Lambda_0\right)}$.
These equations have been checked against the known limits: the 2HDM \cite{Branco:2011iw}; the MW model \cite{He:2013tla}.

When we include the effect of the coupling $g'$ in the RGE we end up with high scale couplings that no longer satisfy custodial symmetry. The deviations from the symmetry limit are small as expected, proportional to $g'$, and we ignore them in our numerical analysis.

\subsection{The running of the scalar couplings}

As we run the couplings between the electroweak and high scales we find three possible behaviors: well behaved couplings at all scales; one or more of the couplings develop a LP; or even though there are no LP for the scales considered, perturbative unitarity or stability are violated at some scale in the range. We illustrate these three possibilities below. 

\begin{figure}[thb]
\includegraphics[width=.45\textwidth]{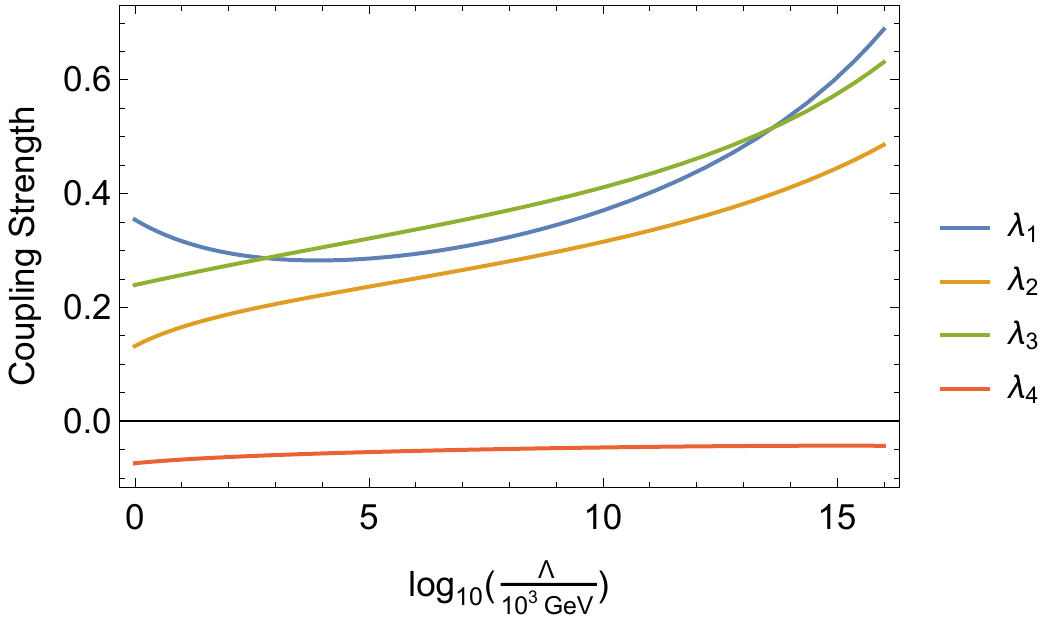}
\includegraphics[width=.45\textwidth]{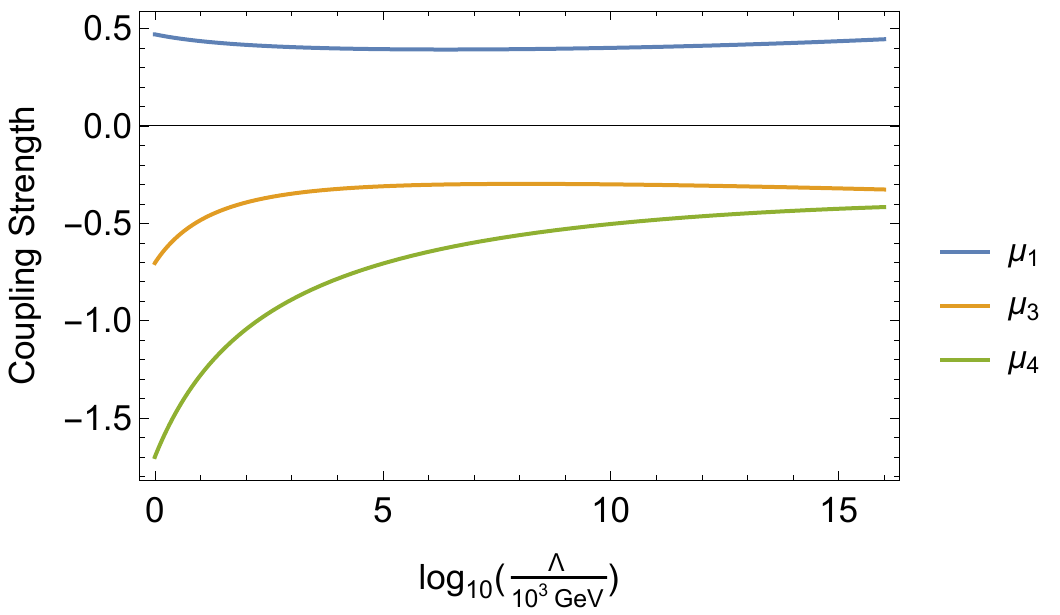}
\includegraphics[width=.45\textwidth]{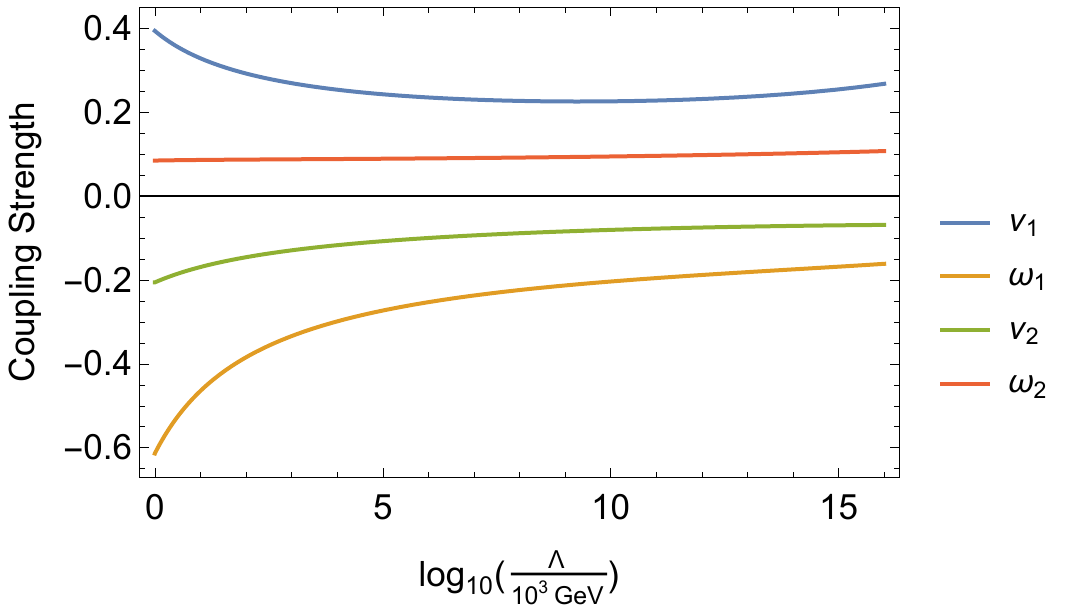}
\includegraphics[width=.45\textwidth]{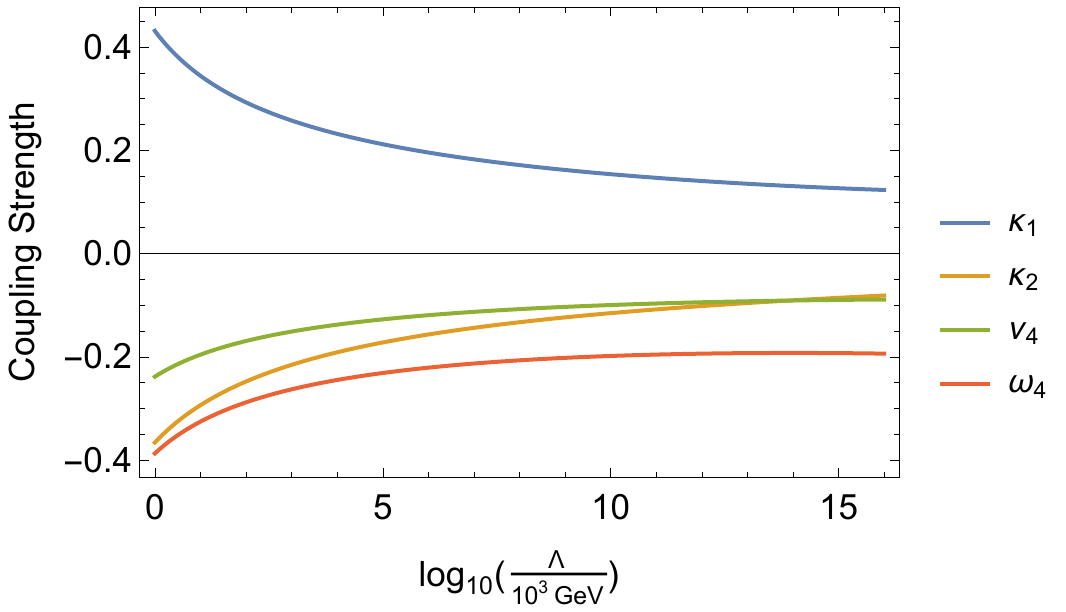}
\caption{Running couplings for a case that satisfies unitarity and stability conditions at all scales below $\Lambda_{\rm Planck}$. }
\label{f:running1}
\end{figure}

\begin{figure}[thb]
\includegraphics[width=.45\textwidth]{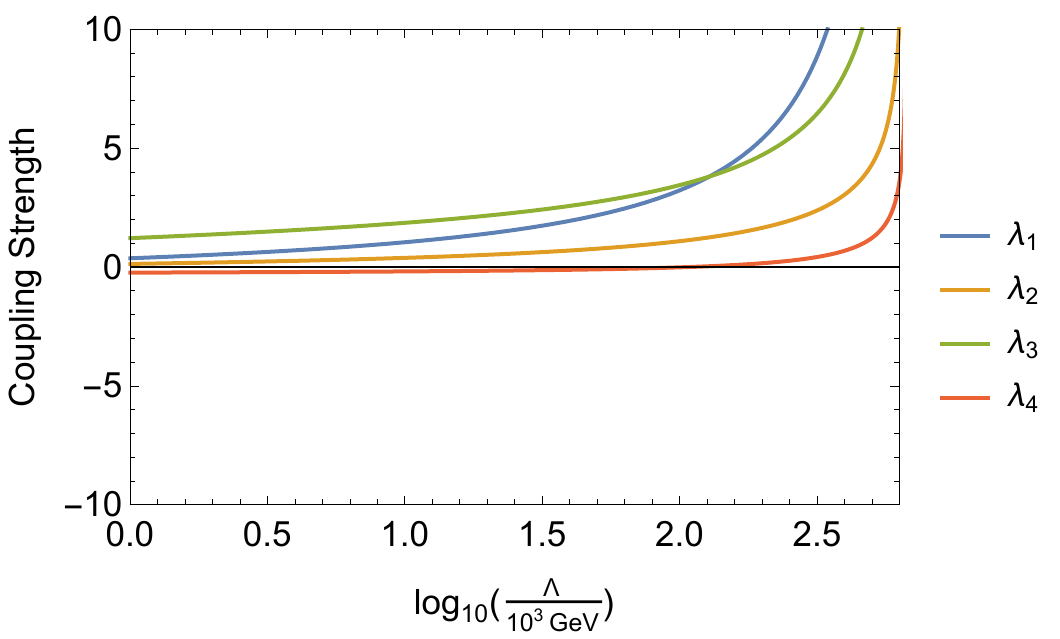}
\includegraphics[width=.45\textwidth]{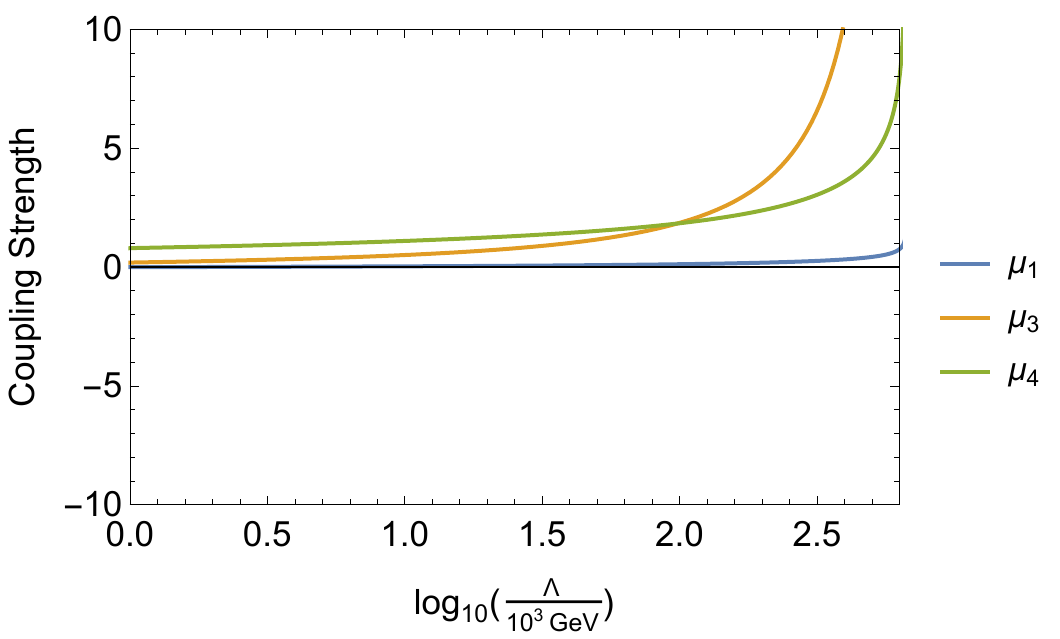}
\includegraphics[width=.45\textwidth]{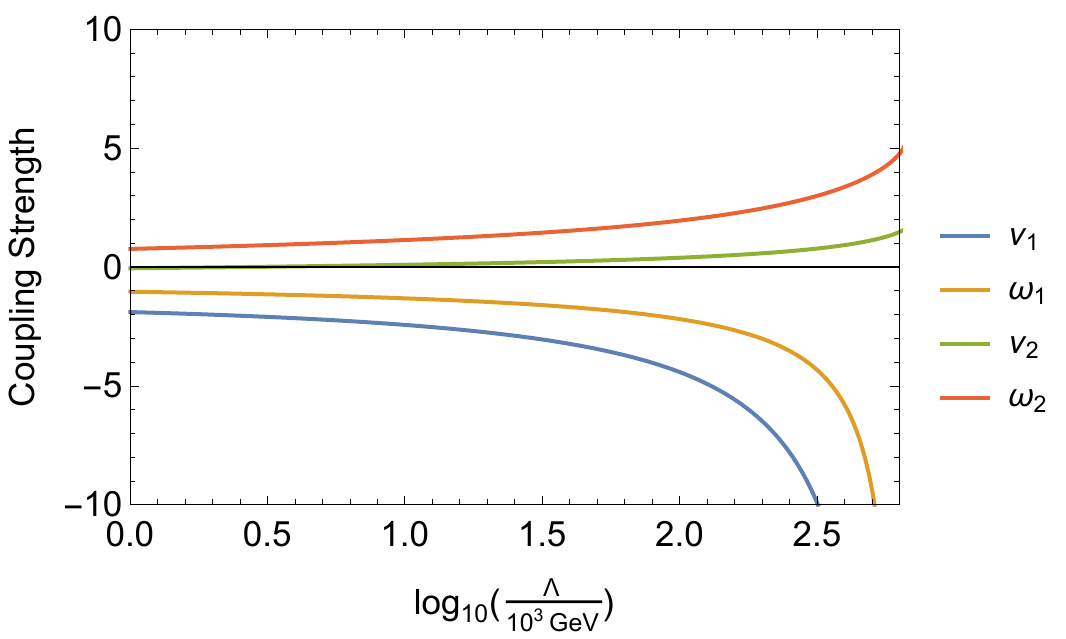}
\includegraphics[width=.45\textwidth]{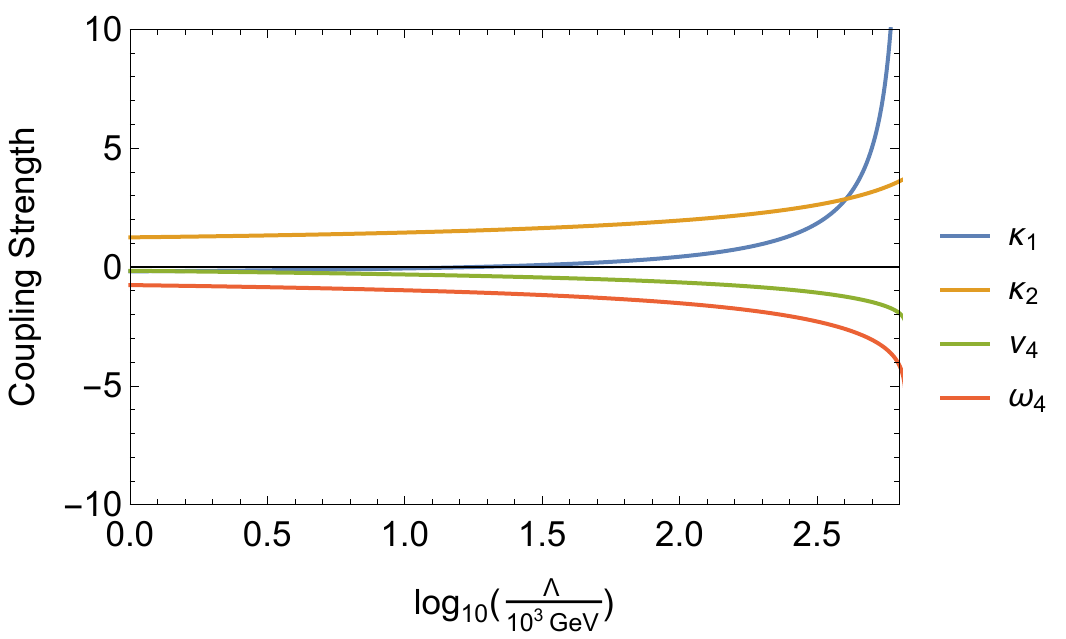}
\caption{Running couplings for a case where a LP is encountered below  $\Lambda_{\rm Planck}$. }
\label{f:running2}
\end{figure}

\begin{figure}[thb]
\includegraphics[width=.45\textwidth]{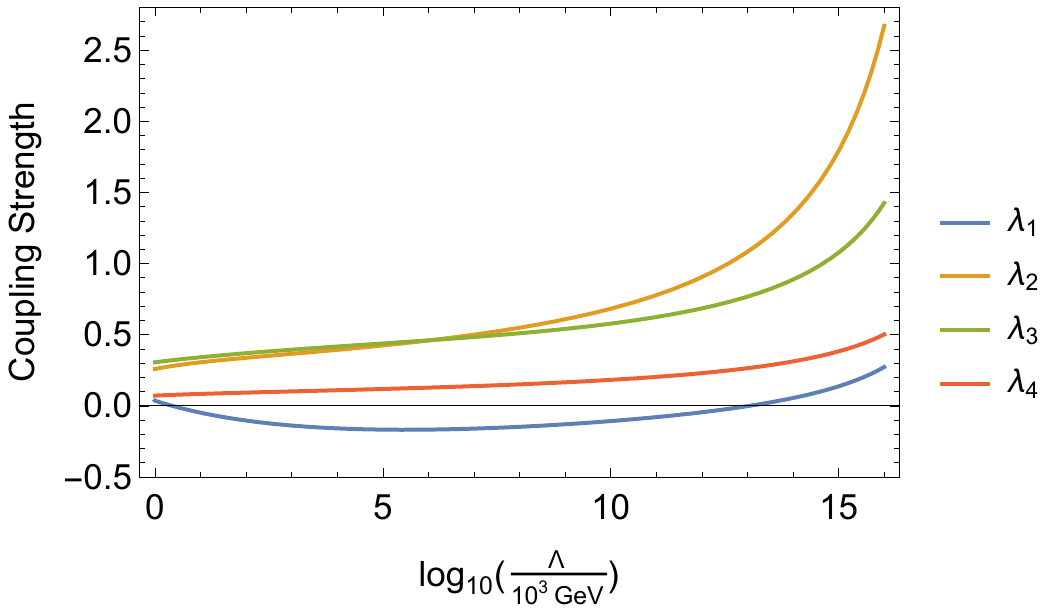}
\includegraphics[width=.45\textwidth]{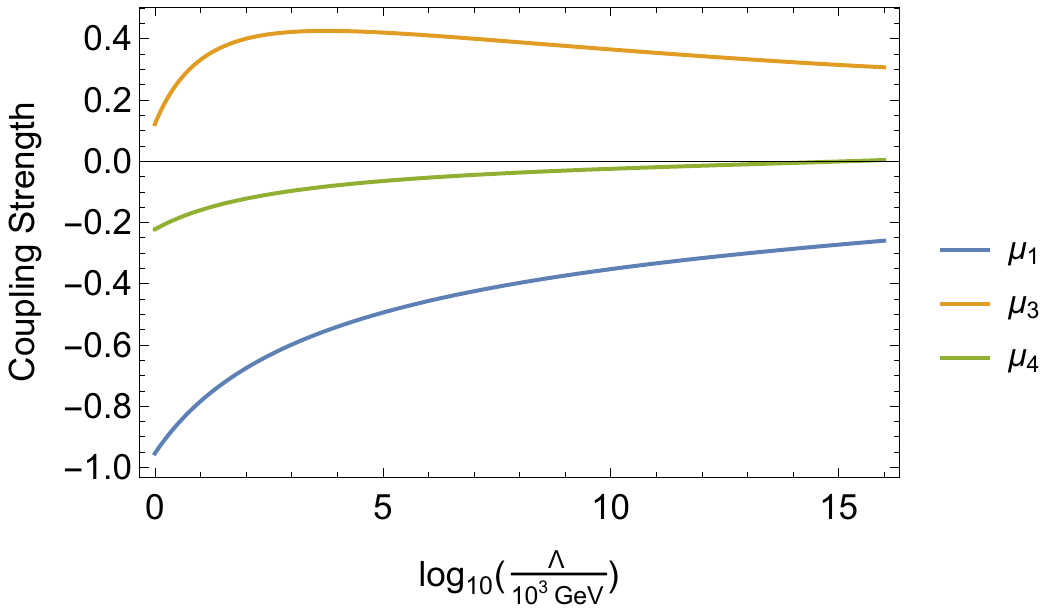}
\includegraphics[width=.45\textwidth]{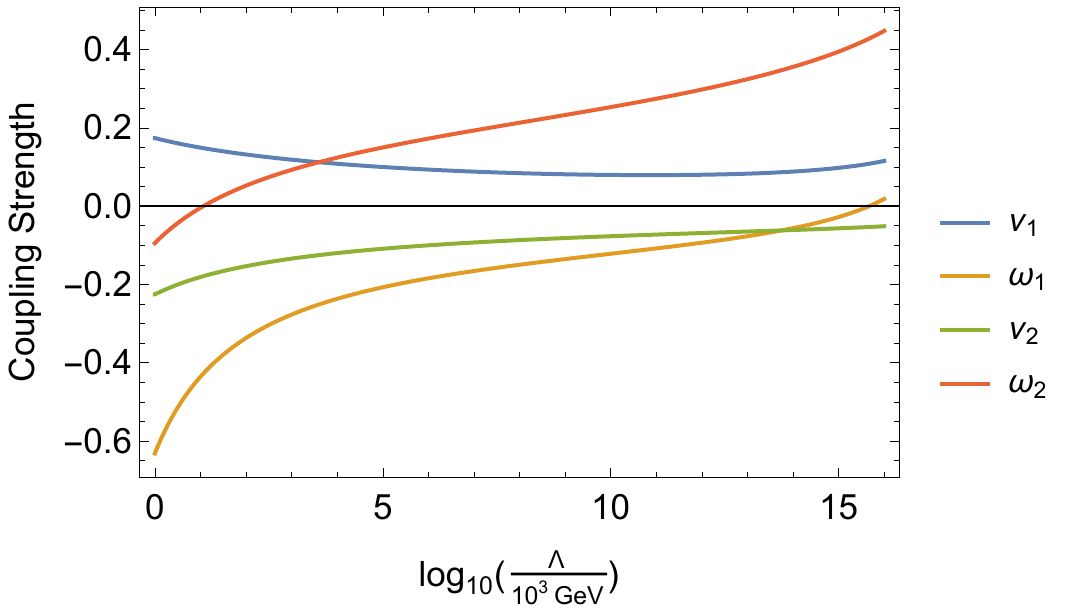}
\includegraphics[width=.45\textwidth]{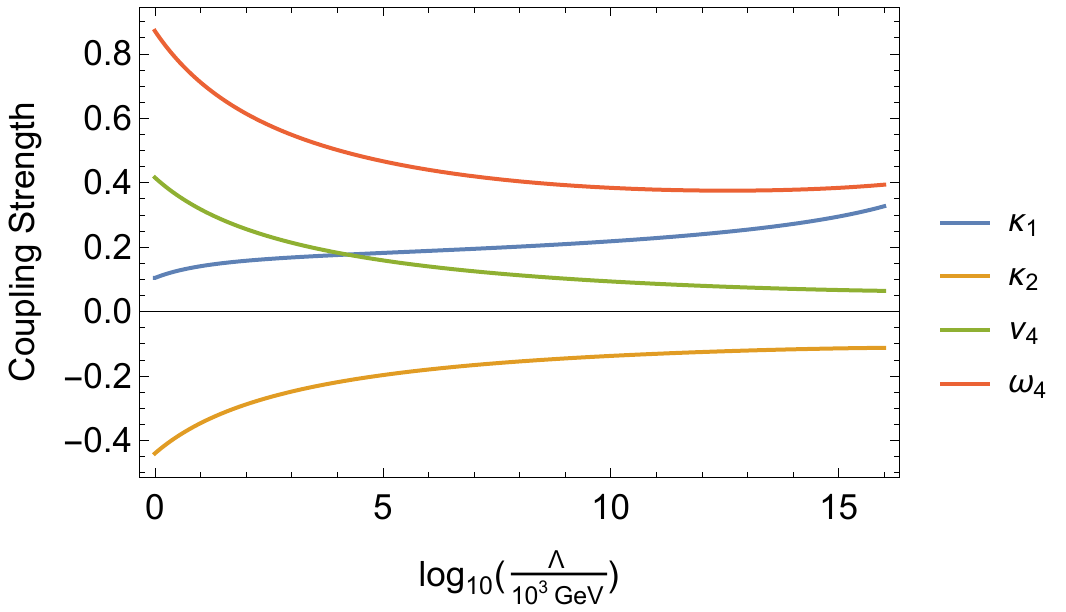}
\caption{Running couplings for a case where no LP is encountered below $\Lambda_{\rm Planck}$, but the unitarity and stability conditions are not satisfied for all scales $\Lambda < \Lambda_{\rm Planck}$. }
\label{f:running3}
\end{figure}

Figure \ref{f:running1} illustrates the case of well behaved couplings up to the Planck scale. Figure  \ref{f:running2} shows how it is possible to develop multiple LP at relatively low energy scales even when the couplings are small and perturbative at the electroweak scale. Finally, figure \ref{f:running3} illustrates a case where there are no LP below the Planck scale, but perturbative unitarity is violated  at some point below $\Lambda_{\rm Planck}$. In this case, the violation is due to the presence of a LP just beyond $\Lambda_{\rm Planck}$.

\subsection{Constraining the allowed region of parameter space}

For this purpose we use Mathematica to solve the RGE numerically with initial conditions at the electroweak scale. The initial conditions used for the new couplings are points that satisfy perturbative unitarity and stability determined as in Ref.~\cite{Cheng:2016tlc}. For each point we evolve all the couplings up to a high scale $\Lambda$ and discard the point if a LP is detected, or if the couplings at any scale below $\Lambda$ violate the perturbative unitarity or stability conditions. This results in acceptable points satisfying a more stringent condition than the absence of LP, in the spirit of renormalization group improved unitarity bounds of Ref.~\cite{Marciano:1989ns}. The use of this condition in our numerical search makes it easier to find acceptable points than if we were to allow a LP at $\Lambda$. It would also be possible to constrain the parameters using higher order unitarity conditions, but we do not pursue this in this paper \cite{Cacchio:2016qyh,Murphy:2017ojk}.

The parameter space is too large for a completely random scan to be efficient. Instead, we follow the approach described below.

\begin{enumerate}

\item We begin our study at an intermediate energy scale $\Lambda_m$ which, for the sake of computational efficiency. We choose (by trial and error) it to be $\ln (\Lambda_m/{10^3\ {\rm GeV}}) = 10$.

\item Before running the RGE for the whole model, we generate a large sample of points within the 2HDM subspace. The sample is generated in such a way that a large portion of it is valid up to $\Lambda_m$.

\item We then use this 2HDM data set as seeds, randomly assigning values to the new couplings within a proper range, to generate a sample for the whole parameter space. From this sample we find a few hundred valid points and determine the hypercube which contains most of the solutions, a region somewhat smaller than that allowed by perturbative unitarity at the electroweak scale.

\item Starting from these few hundred points we study nearby points to expand the allowed region.

\item We finally construct the region of parameter space where the full model is valid up to the scale $\Lambda_m$  by repeating step 4 recursively for a sufficiently long time.

\item Points that are valid up to scales higher than $\Lambda_m$, are generally inside a sub-region of the allowed region up to scale $\Lambda_m$. Therefore, to find the constraints for a higher scale, we select the seed points from step 5 and repeat step 4 to construct the new allowed region.

\end{enumerate}

\begin{figure}[thb]
\includegraphics[width=.45\textwidth]{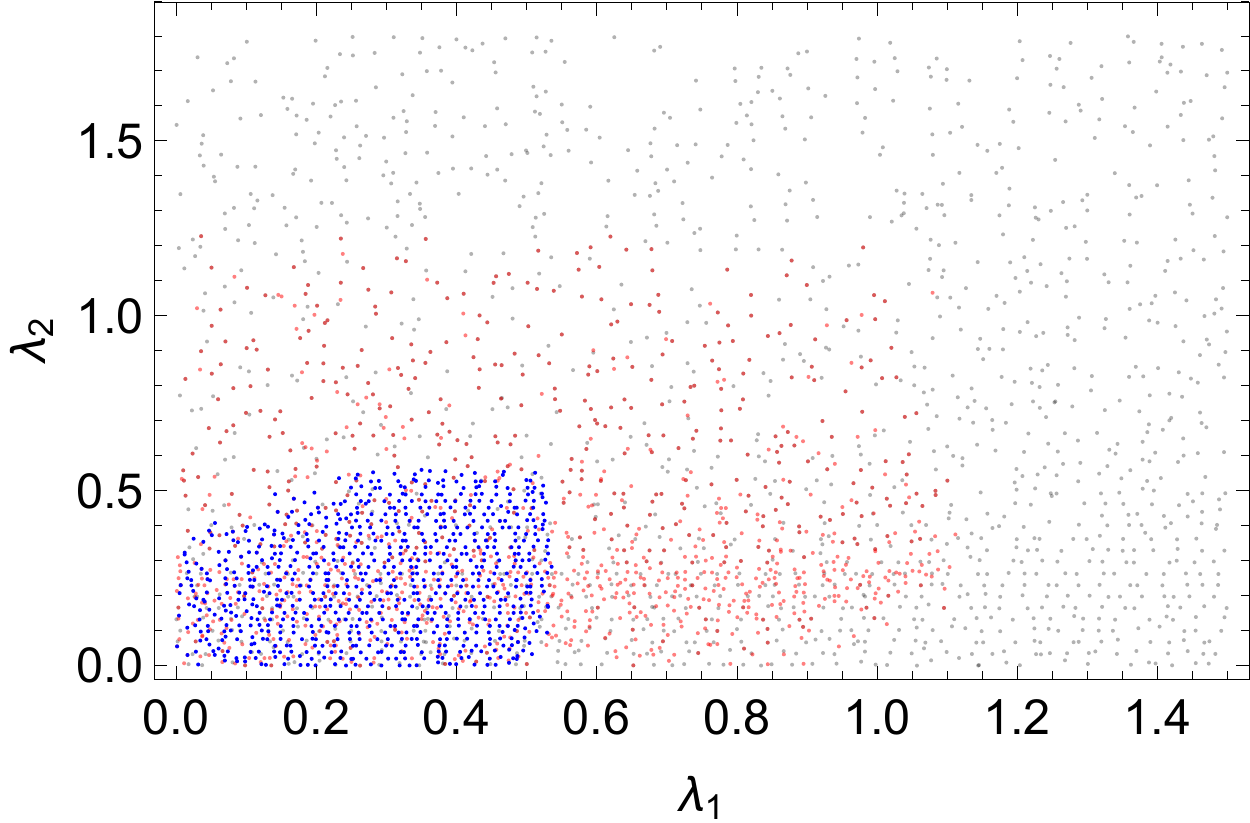}
\includegraphics[width=.45\textwidth]{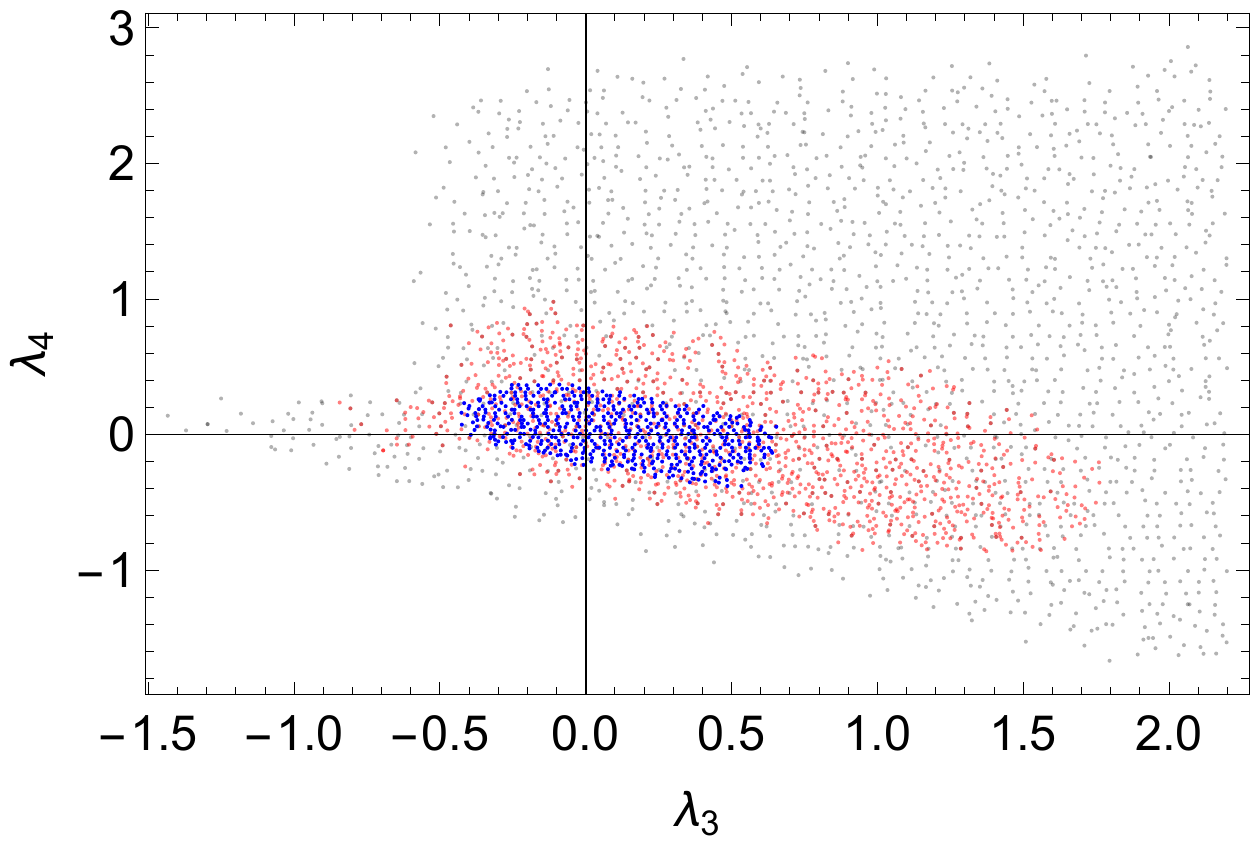}
\includegraphics[width=.45\textwidth]{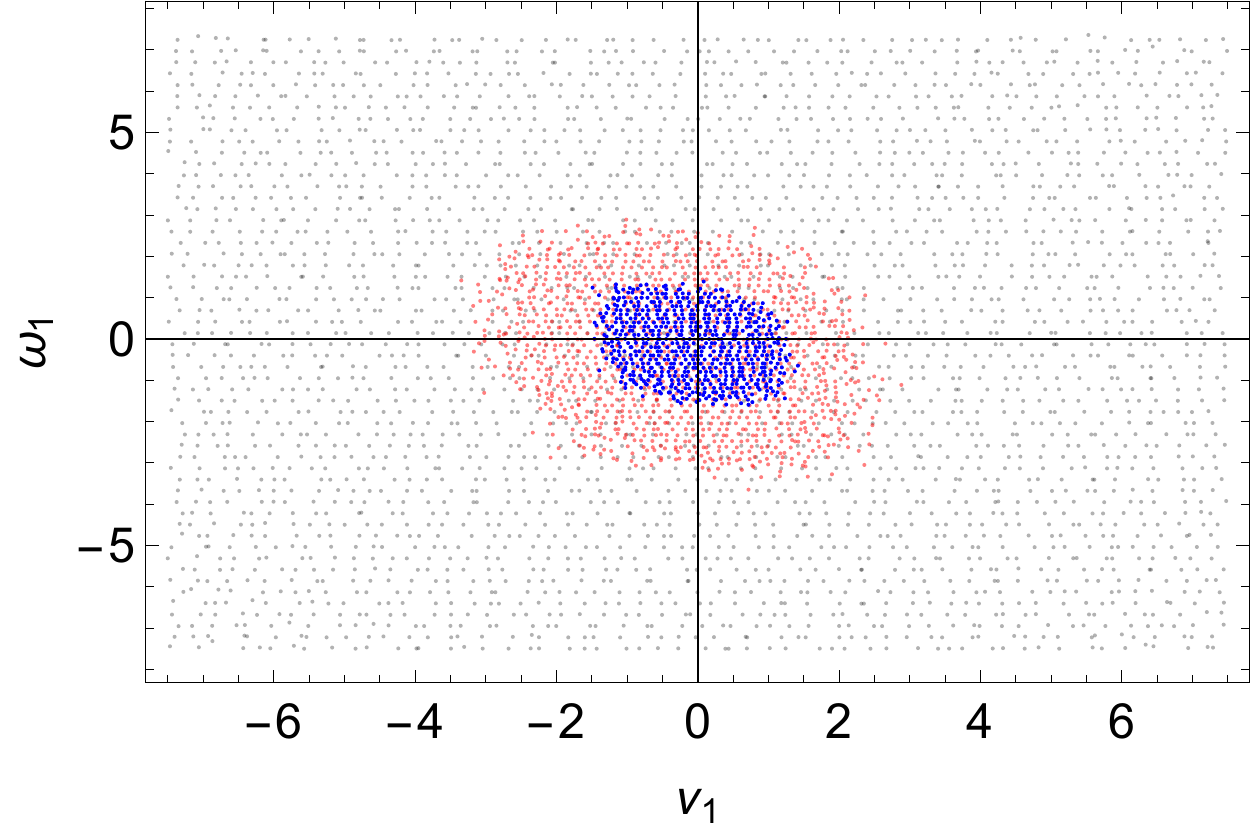}
\includegraphics[width=.45\textwidth]{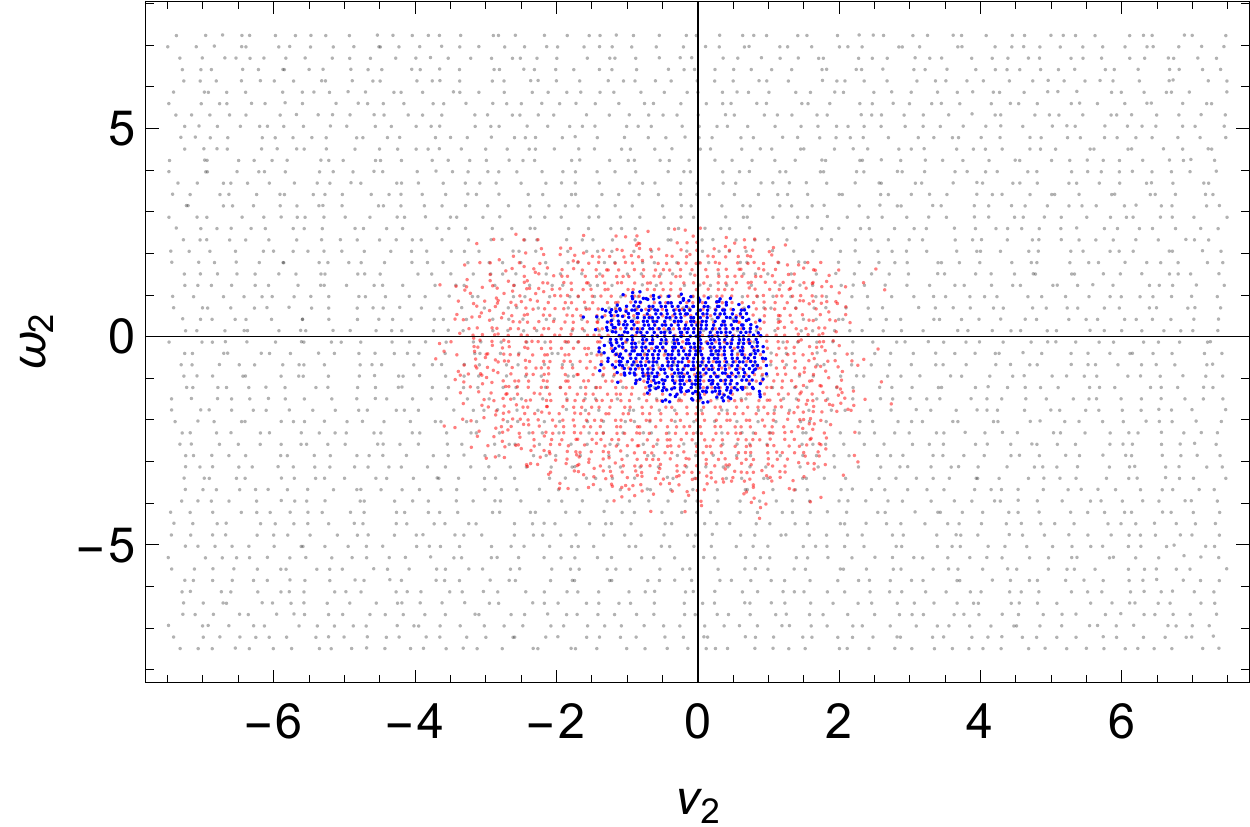}
\includegraphics[width=.45\textwidth]{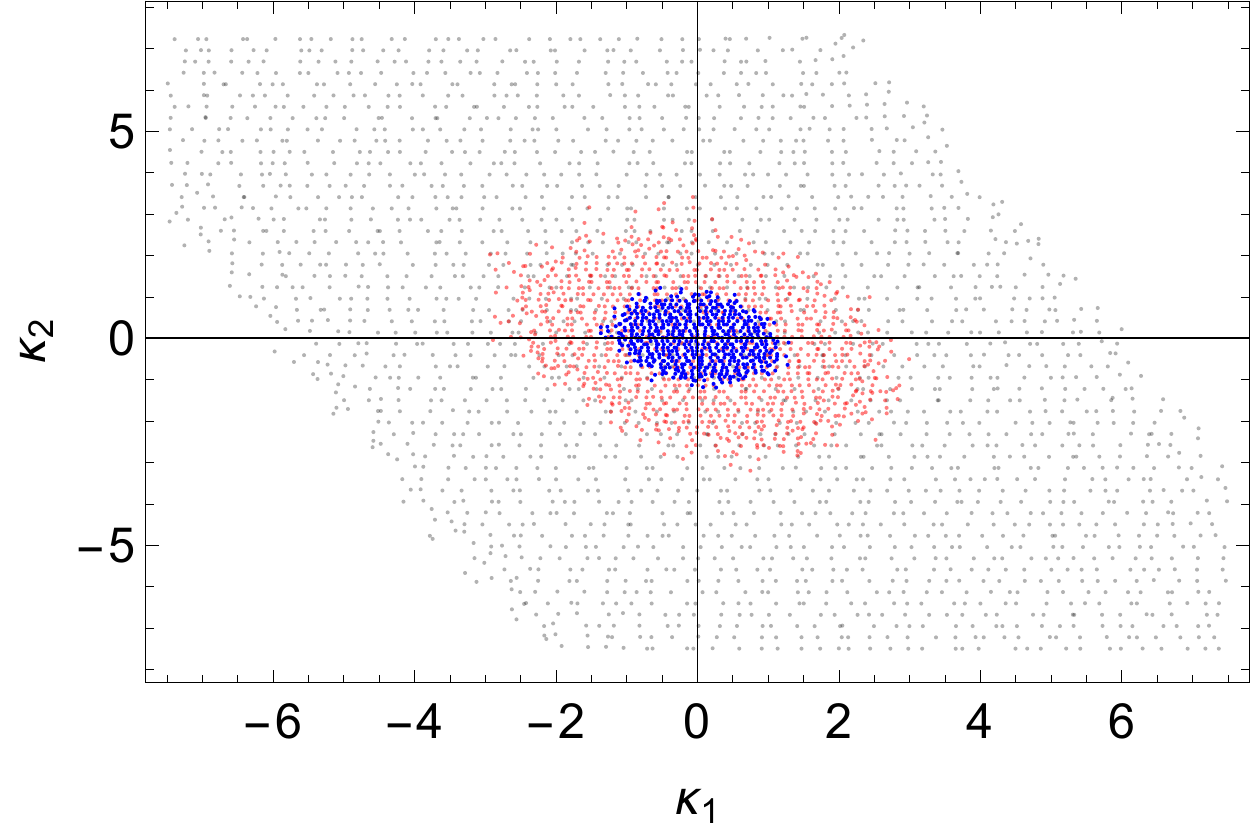}
\includegraphics[width=.45\textwidth]{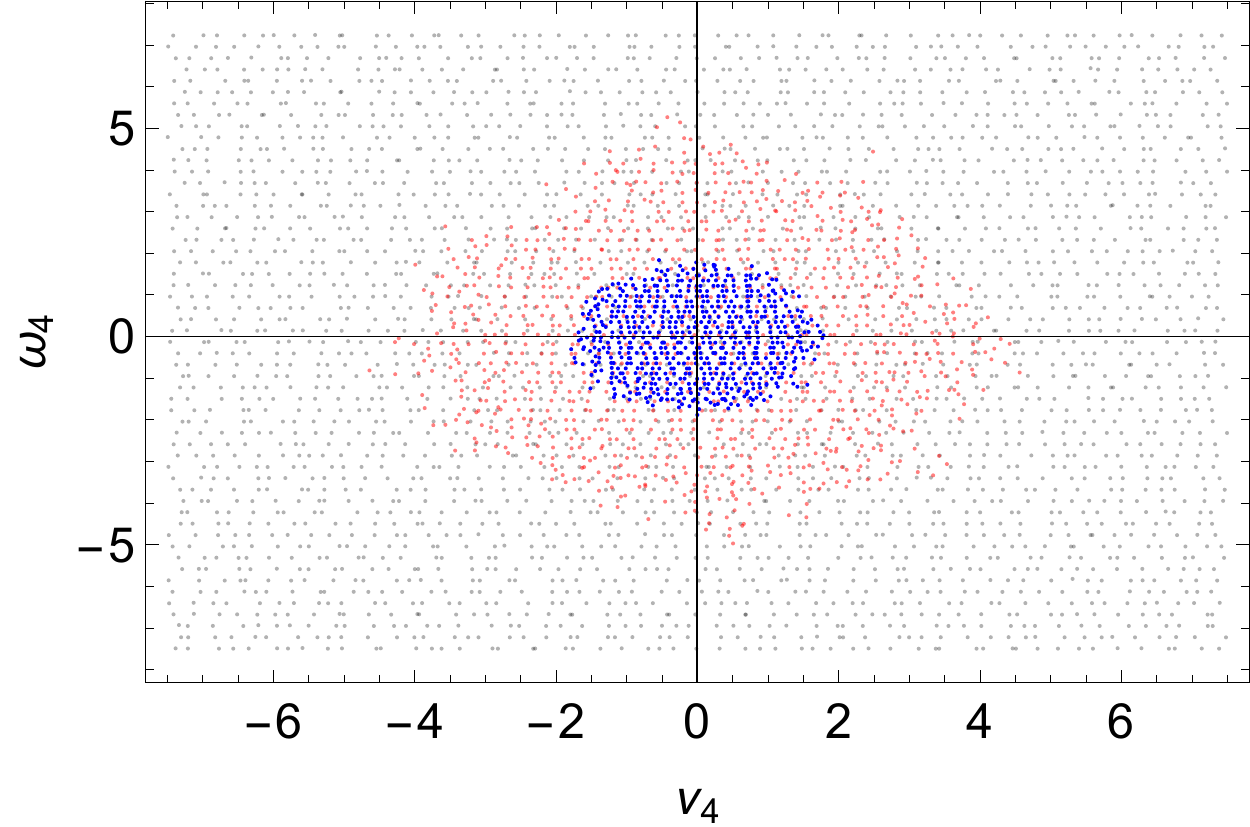}
\includegraphics[width=.45\textwidth]{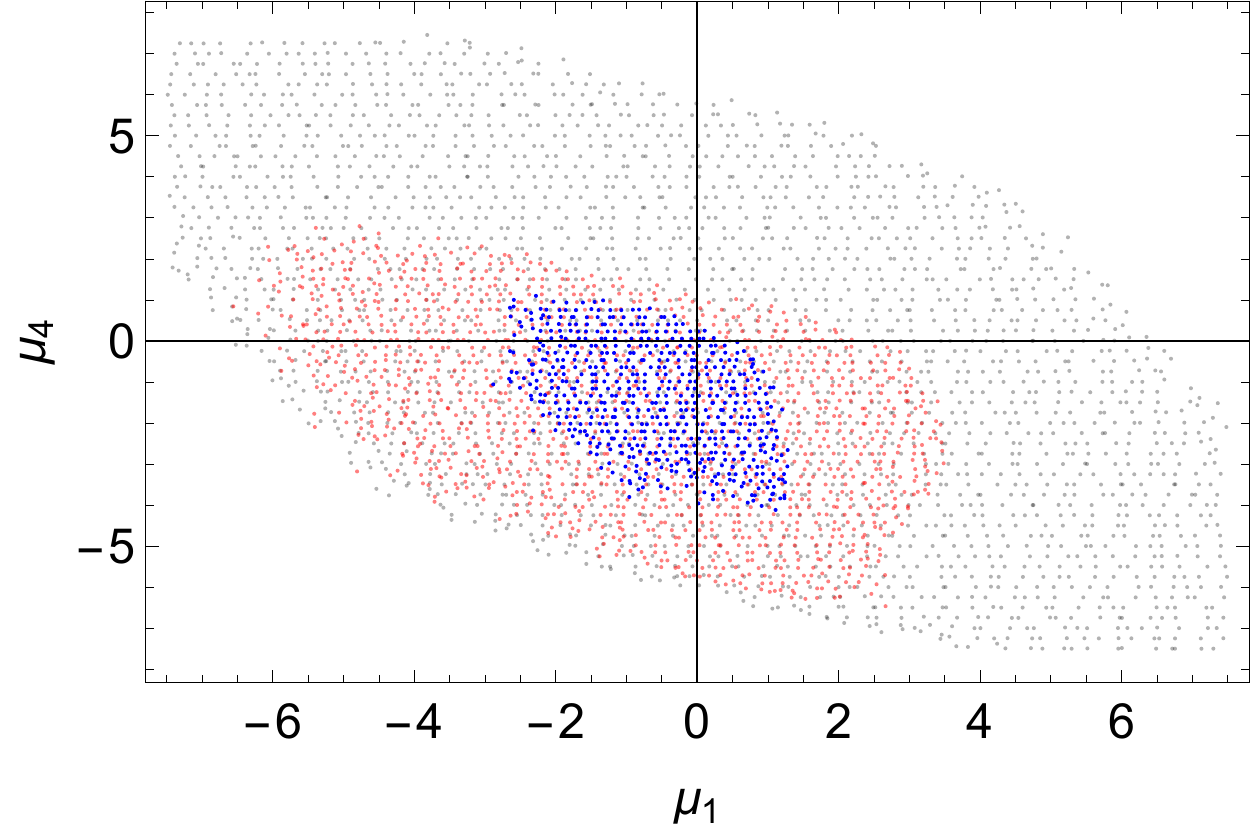}
\includegraphics[width=.45\textwidth]{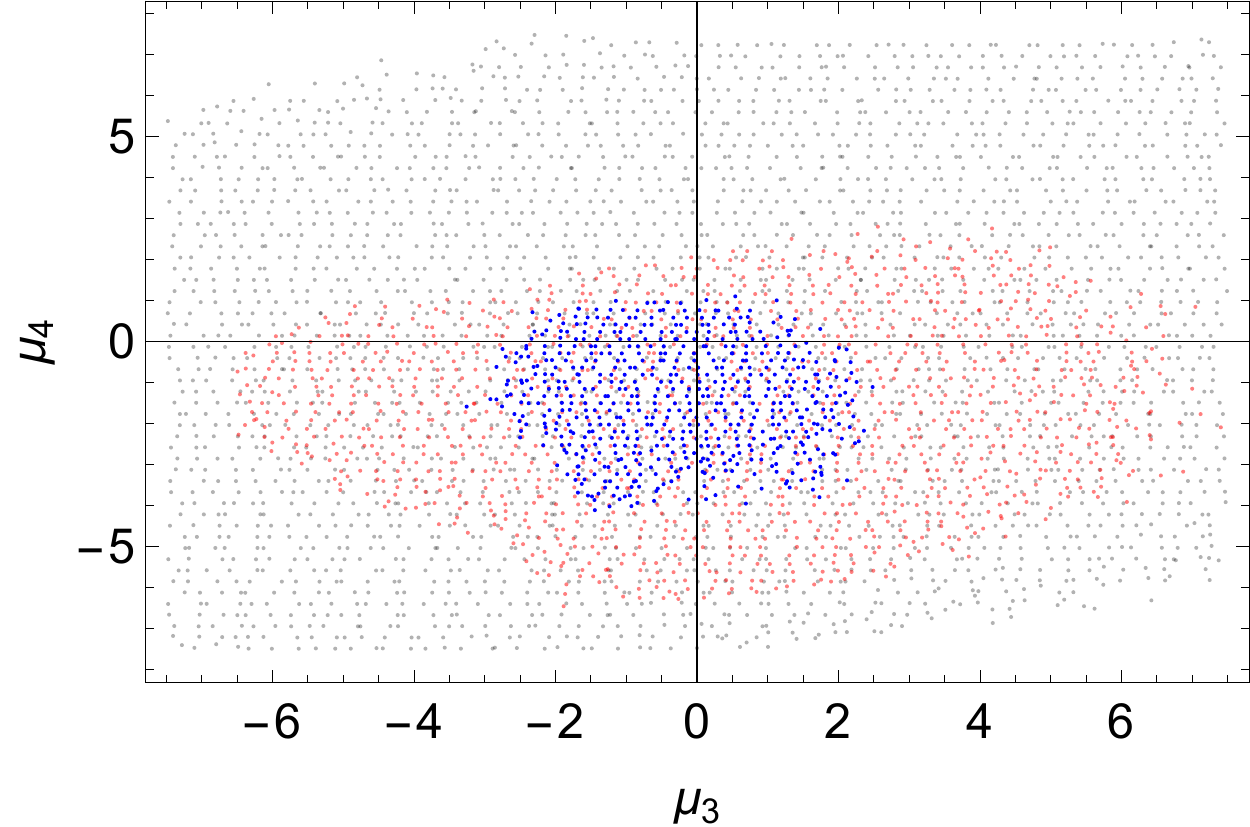}
\caption{Representative two-dimensional projections of the allowed parameter space for which the model is valid up to the intermediate scale $\Lambda_m$ (red) and $\Lambda_{\rm GUT}$ (blue). The black points are the seed points used as initial values at the electroweak scale that satisfy both perturbative unitarity and stability.
\label{f:uni-GUT}}
\end{figure}

Our results are illustrated in Figure~\ref{f:uni-GUT} for the scales $\Lambda_m$ and $\Lambda_{\rm GUT}$ in representative two-dimensional projections. The GUT scale is chosen because of the existing $SU(5)$ \cite{Perez:2016qbo} or  $SO(10)$ \cite{Bertolini:2013vta} models which can have TeV scale scalar color octets, but the figures illustrate the general trend as we require the model to be valid up to higher energy scales. In general, for the new parameters involving the color-octet scalars, the allowed parameter space is now very significantly reduced with respect to that allowed by tree-level unitarity. In addition, this procedure produces the first constraints on parameters like $\nu_4$ and $\omega_4$ which do not affect two to two processes at tree-level.

Validity up to the GUT scale thus results in approximate one at a time constraints
\begin{eqnarray}
0\leq \lambda_{1,2} \lsim 0.5, && -0.4 \lsim \lambda_3 \lsim 0.7,\nonumber \\
-0.4 \leq \lambda_{4} \lsim 0.4, && |\omega_{1,2}| \lsim 1.6, \nonumber \\
|\nu_{1,2}| \lsim 1.6, && |\kappa_{1,2}| \lsim 1.4, \nonumber \\
|\nu_{4}| \lsim 2, && |\omega_{4}| \lsim 2, \nonumber \\
|\mu_{1,3}| \lsim 3.3, && |\mu_4| \lsim 4.2.
\end{eqnarray}

\begin{figure}[thb]
\includegraphics[width=.45\textwidth]{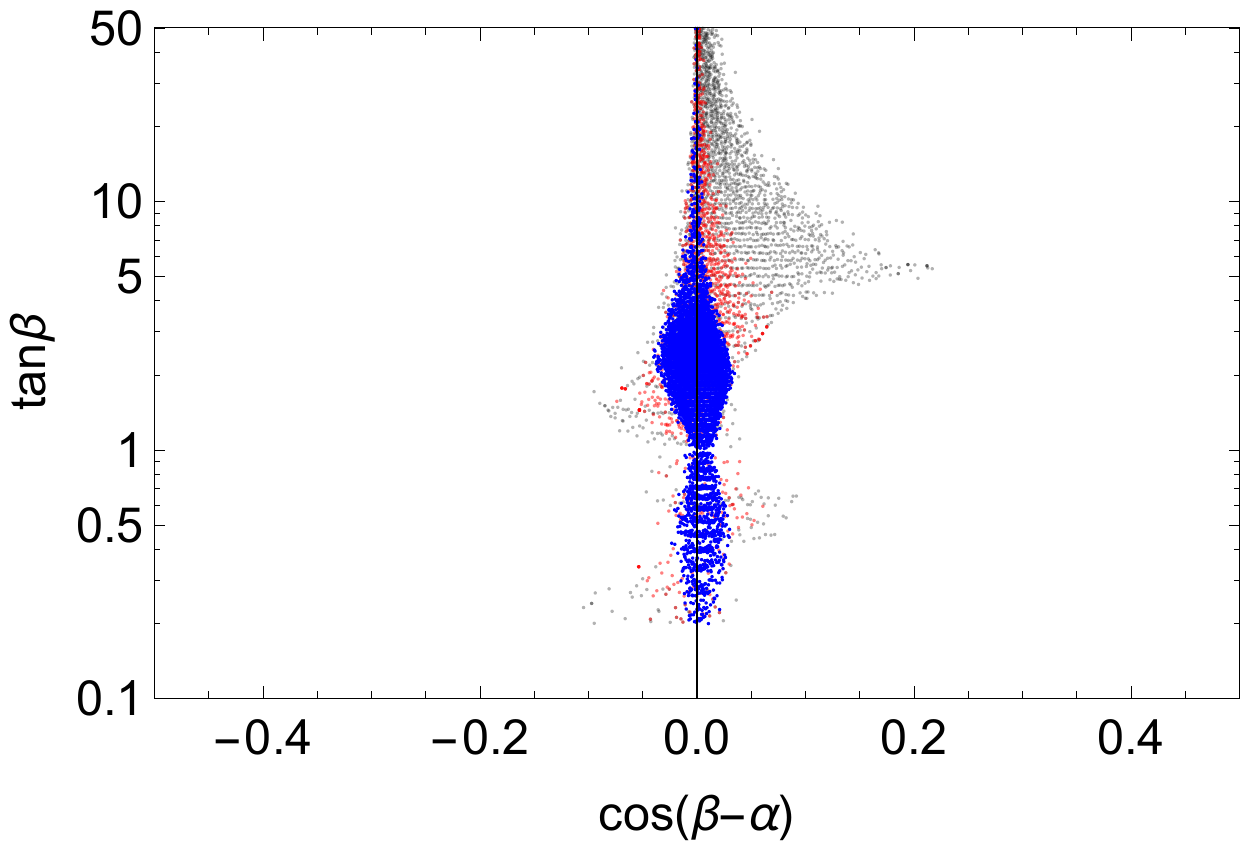}
\includegraphics[width=.45\textwidth]{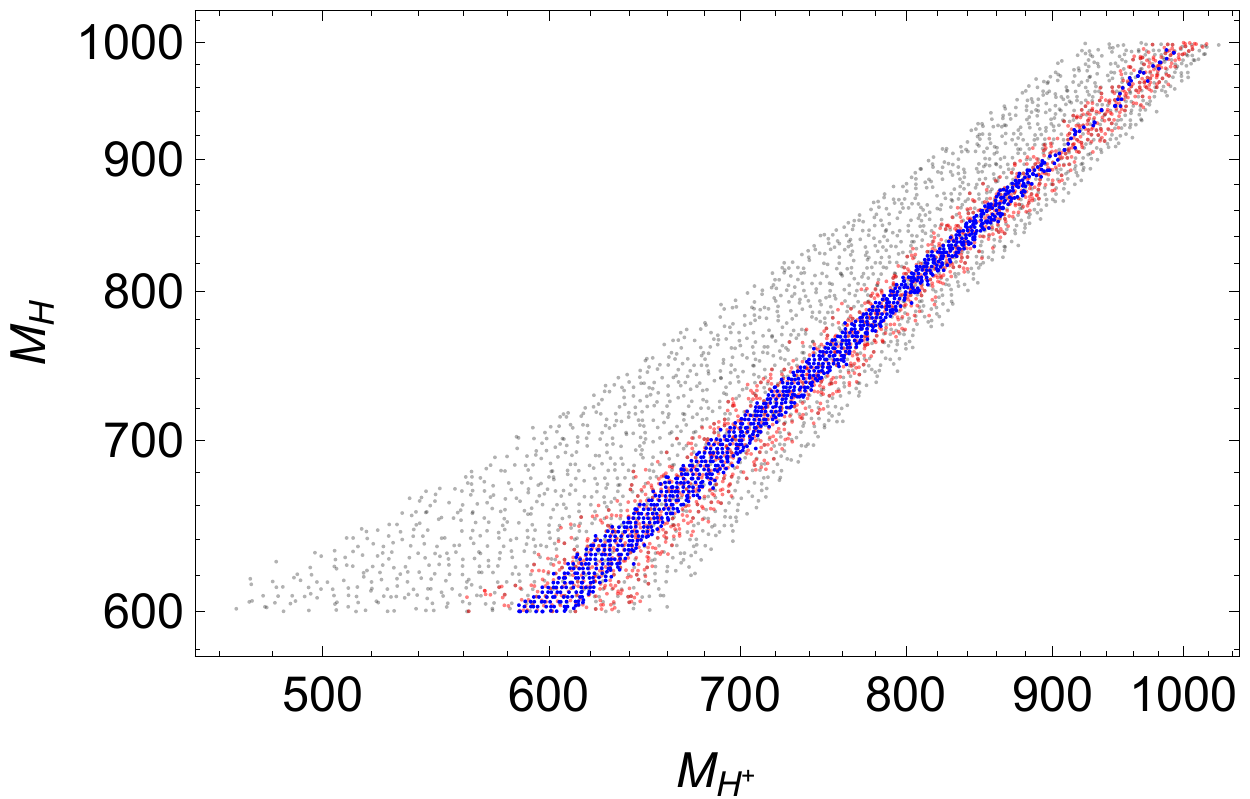}
\caption{Points for which the 2HDM model satisfies both perturbative unitarity and stability at the electroweak scale (black) compared to those for which it is also valid up to the scale $\Lambda_m$ (red) and $\Lambda_{\rm GUT}$ (blue).
\label{f:2hdm-GUT}}
\end{figure}

For the parameters of the 2HDM the points that produce a model valid to high energy scales are those for which $\cos(\beta-\alpha)$ is closer to zero, as in the alignment limit; and those for which $M_H$ is very close to $M_{H^\pm}$ as shown in Figure~\ref{f:2hdm-GUT}. 

\section{Conclusions}

In this paper, we provide the renormalization group equations for all the couplings in the scalar potential of a 2HDM augmented with a color-octet. As an application, we constrain the parameters of the model by requiring it to be valid up to some high scale. 
The acceptable region of the parameter space that satisfies both unitarity and stability constraints without developing LP up to a high scale is determined numerically, and the resulting constraints are provided for the case $\Lambda_{\rm HIGH}=\Lambda_{\rm GUT}$.  As expected, the allowed region is reduced as the scale increases; looking at the 2HDM subspace, it contracts towards the alignment limit and mass-degeneracy of heavy neutral and charged Higgses.

\begin{acknowledgments}

We thank Dianne Cook for help visualizing the multidimensional parameter space with the aid of the grand and guided tour \cite{method} with the software GGOBI \cite{visual}.
Li Cheng thanks David Atwood and Kerry Whisnant for useful discussions. The work of GV was supported in part by the  Australian Government through the Australian Research Council.

\end{acknowledgments}

\pagebreak

\appendix

\section{General renormalization group equations}

In this appendix we collect the general results for the RGE, without the assumptions of custodial and CP symmetry used for our numerical study.

\begingroup
\allowdisplaybreaks
\begin{align*}
&16\pi^2\beta_{\lambda_1} = 12\lambda_1^2 + 4\lambda_3^2 + 4\lambda_3\lambda_4 + 2\lambda_4^2 + 2\lambda_5^2
+ 8\nu_1^2 + 8\nu_1\nu_2 + 4\nu_2^2 + 16\nu_3^2\\
&\qquad\qquad\quad - 12 \lambda_t^4 - 3\lambda_1\left( 3g^2 - 4\lambda_t^2 + g'^2 \right) + \frac34 \left( 3g^4 + 2g^2 g'^2 + g'^4 \right),\\
&16\pi^2\beta_{\lambda_2} = 12\lambda_2^2 + 4\lambda_3^2 + 4\lambda_3\lambda_4 + 2\lambda_4^2 + 2\lambda_5^2
+ 8\omega_1^2 + 8\omega_1\nu_2 + 4\omega_2^2 + 16\omega_3^2\\
&\qquad\qquad\quad - 3\lambda_2\left( 3g^2 + g'^2 \right) + \frac34 \left( 3g^4 + 2g^2 g'^2 + g'^4 \right),\\
&16\pi^2\beta_{\lambda_3} = 4\lambda_3^2 + 2\lambda_4^2 + 2 \lambda_5^2 + 2\left(\lambda_1+\lambda_2\right)\left(3\lambda_3 + \lambda_4\right)
+ 8\nu_1\omega_1 + 4\nu_1\omega_2 + 4\nu_2\omega_1 \\
&\qquad\qquad\quad  + 4\left|\kappa_2\right|^2 + 4\left|\kappa_3\right|^2 - 3\lambda_3\left( 3g^2 - 2\lambda_t^2 + g'^2 \right) + \frac34 \left( 3g^4 - 2g^2 g'^2 + g'^4 \right),\\
&16\pi^2\beta_{\lambda_4} = 4\lambda_4^2 + 8\lambda_3\lambda_4 + 8\lambda_5^2 + 2\left(\lambda_1+\lambda_2\right)\lambda_4 + 4\nu_2\omega_2
+ 8\left|\kappa_1\right|^2 + 4\kappa_1\kappa_2^* + 4\kappa_1^*\kappa_2\\
&\qquad\qquad\quad  + 4\left|\kappa_3\right|^2 + 3g^2g'^2 - 3\lambda_4\left(3g^2 - 2\lambda_t^2 + g'^2\right),\\
&16\pi^2\beta_{\lambda_5} = 2\left(\lambda_1 + \lambda_2 + 4\lambda_3 + 6\lambda_4\right) \lambda_5
+8\kappa_1^2+8\kappa_1\kappa_2+4\kappa_2^2+16\nu_3\omega_3\\
&\qquad\qquad\quad - 3\lambda_5\left(3g^2 - 2\lambda_t^2 + g'^2\right),\\
&16\pi^2\beta_{\nu_1} = 6\lambda_1\nu_1 + 2\lambda_1\nu_2 + 4\lambda_3\omega_1 + 2\lambda_3\omega_2 + 2\lambda_4\omega_1 + 2\nu_1^2 + \nu_2^2 + 4\nu_3^2 \\
&\qquad\qquad\quad + 2\left|\kappa_1\right|^2 + \left|\kappa_2\right|^2 + \left|\kappa_3\right|^2 + \nu_1\left(8\mu_1 + 8 \mu_2 + 17\mu_3 + 10\mu_4 + 3\mu_5 + 5\mu_6\right)
 \\
&\qquad\qquad\quad
+\nu_2\left(\frac83\mu_1 + \frac83\mu_2 + 8\mu_3 + \mu_4 + \mu_5 + \frac83\mu_6\right) \\ 
&\qquad\qquad\quad - \frac23\left|\nu_4\right|^2 + \frac7{3}\nu_4\nu_5^* + \frac7{3}\nu_4^*\nu_5 - \frac23\left|\nu_5\right|^2,\\
&16\pi^2\beta_{\nu_2} = 2\lambda_1\nu_2 + 2\lambda_4\omega_2 + 4\nu_1\nu_2 + 2\nu_2^2 + 16\nu_3^2  + 2\kappa_1\kappa_2^* + 2\kappa_1^*\kappa_2 + 2\left|\kappa_2\right|^2 + 4\left|\kappa_3\right|^2 \\
&\qquad\qquad\quad + \nu_2\left(\frac83\mu_1 + \frac83\mu_2 + \mu_3 + 8\mu_4 + \mu_5 - \frac13\mu_6\right) \\
&\qquad\qquad\quad + \frac{17}6\left|\nu_4\right|^2 + \frac4{3}\nu_4\nu_5^* + \frac4{3}\nu_4^*\nu_5 + \frac{17}6\left|\nu_5\right|^2,\\
&16\pi^2\beta_{\nu_3} = 4\nu_1\nu_3 + 6\nu_2\nu_3 + 2\lambda_1\nu_3 + 2\lambda_5\omega_3 + 2\kappa_1\kappa_3 + 3\kappa_2\kappa_3 \\
&\qquad\qquad\quad + \nu_3\left(-\frac13\mu_1 - \frac13\mu_2 + \mu_3 + \mu_4 + 8\mu_5 + \frac83\mu_6\right) \\
&\qquad\qquad\quad + \frac{17}{12}\nu_4^2 + \frac43 \nu_4\nu_5 + \frac{17}{12}\nu_5^2,\\
&16\pi^2\beta_{\nu_4} = 2\nu_3\nu_4^* + 8\nu_3\nu_5^*  + \kappa_3\omega_4^* + 4\kappa_3\omega_5^* 
+ \left(3\kappa_1 + 2\kappa_2\right)\omega_4 + 2\kappa_2\omega_5\\
&\qquad\qquad\quad + \nu_4\left(3\nu_1 + 2\nu_2 + 6\mu_1 + 2\mu_2 + 3\mu_3 + 2\mu_4 + \mu_5 + \mu_6\right)\\
&\qquad\qquad\quad + \nu_5\left(2\nu_2  - \mu_2 + 2\mu_4 + 4\mu_5 + \mu_6\right),\\
&16\pi^2\beta_{\nu_5} = 8\nu_3\nu_4^* + 2\nu_3\nu_5^* + 4\kappa_3\omega_4^* + \kappa_3\omega_5^*  
+ 2\kappa_2\omega_4 + \left(3\kappa_1 + 2\kappa_2\right)\omega_5\\
&\qquad\qquad\quad + \nu_4\left(2\nu_2  - \mu_1 + 2\mu_4 + 4\mu_5 + \mu_6\right)\\
&\qquad\qquad\quad + \nu_5\left(3\nu_1 + 2\nu_2 + 6\mu_1 + 2\mu_2 + 3\mu_3 + 2\mu_4 + \mu_5 + \mu_6\right),\\
&16\pi^2\beta_{\omega_1} = 6\lambda_2\omega_1 + 2\lambda_2\omega_2 + 4\lambda_3\nu_1 + 2\lambda_3\nu_2 + 2\lambda_4\nu_1 + 2\omega_1^2 + \omega_2^2 + 4\omega_3^2\\
&\qquad\qquad\quad + 2\left|\kappa_1\right|^2 + \left|\kappa_2\right|^2 + \left|\kappa_3\right|^2 + \omega_1\left(8\mu_1 + 8 \mu_2 + 17\mu_3 + 10\mu_4 + 3\mu_5 + 5\mu_6\right) \\
&\qquad\qquad\quad
+\omega_2\left(\frac83\mu_1 + \frac83\mu_2 + 8\mu_3 + \mu_4 + \mu_5 + \frac83\mu_6\right) \\
&\qquad\qquad\quad - \frac23\left|\omega_4\right|^2 + \frac7{3}\omega_4\omega_5^* + \frac7{3}\omega_4^*\omega_5 - \frac23\left|\omega_5\right|^2,\\
&16\pi^2\beta_{\omega_2} = 2\lambda_2\omega_2 + 2\lambda_4\nu_2 + 4\omega_1\omega_2 + 2\omega_2^2 + 16\omega_3^2  + 2\kappa_1\kappa_2^* + 2\kappa_1^*\kappa_2 + 2\left|\kappa_2\right|^2 + 4\left|\kappa_3\right|^2 \\
&\qquad\qquad\quad + \omega_2\left(\frac83\mu_1 + \frac83\mu_2 + \mu_3 + 8\mu_4 + \mu_5 - \frac13\mu_6\right) \\
&\qquad\qquad\quad + \frac{17}6\left|\omega_4\right|^2 + \frac4{3}\omega_4\nu_5^* + \frac4{3}\omega_4^*\omega_5 + \frac{17}6\left|\omega_5\right|^2,\\
&16\pi^2\beta_{\omega_3} = 4\omega_1\omega_3 + 6\omega_2\omega_3 + 2\lambda_2\omega_3 + 2\lambda_5\nu_3 + 2\kappa_1^*\kappa_3 + 3\kappa_2^*\kappa_3 \\
&\qquad\qquad\quad + \omega_3\left(-\frac16\mu_1 - \frac16\mu_2 + \frac12\mu_3 + \frac12\mu_4 + 4\mu_5 + \frac43\mu_6\right) \\
&\qquad\qquad\quad + \frac{17}{12}\omega_4^2 + \frac43 \omega_4\omega_5 + \frac{17}{12}\omega_5^2,\\
&16\pi^2\beta_{\omega_4} = 2\omega_3\omega_4^* + 8\omega_3\omega_5^* + \kappa_3\nu_4^* + 4\kappa_3\nu_5^* 
+ \left(3\kappa_1^* + 2\kappa_2^*\right)\nu_4 + 2\kappa_2^*\nu_5\\
&\qquad\qquad\quad + \omega_4\left(3\omega_1 + 2\omega_2 + 6\mu_1 + 2\mu_2 + 3\mu_3 + 2\mu_4 + \mu_5 + \mu_6\right)\\
&\qquad\qquad\quad + \omega_5\left(2\omega_2  - \mu_2 + 2\mu_4 + 4\mu_5 + \mu_6\right),\\
&16\pi^2\beta_{\omega_5} = 8\omega_3\omega_4^* + 2\omega_3\omega_5^* + 4\kappa_3\nu_4^* + \kappa_3\nu_5^*  
+ 2\kappa_2^*\nu_4 + \left(3\kappa_1^* + 2\kappa_2^*\right)\nu_5\\
&\qquad\qquad\quad + \omega_4\left(2\omega_2  - \mu_1 + 2\mu_4 + 4\mu_5 + \mu_6\right)\\
&\qquad\qquad\quad + \omega_5\left(3\omega_1 + 2\omega_2 + 6\mu_1 + 2\mu_2 + 3\mu_3 + 2\mu_4 + \mu_5 + \mu_6\right),\\
&16\pi^2\beta_{\kappa_1} = \kappa_1\left(2\lambda_3 + 4\lambda_4 + 2\nu_1 + 2\omega_1 + 4\mu_1 + 4\mu_2 + 17\mu_3 + 10\mu_4 + 3\mu_5 + 5\mu_6 \right)\\
&\qquad\qquad\quad +\kappa_2\left(2\lambda_4 + \nu_2 + \omega_2 + \frac83\mu_1 + \frac83\mu_2 + 8\mu_3 + 2\mu_4 + 2\mu_5 + \frac83\mu_6\right)\\
&\qquad\qquad\quad + 6\kappa_1^*\lambda_5 + 2\kappa_2^*\lambda_5 + \kappa_3^*\nu_3 + \kappa_3\omega_3
- \frac23\nu_4\omega_4^* +\frac73\nu_4\omega_5^* +\frac73\nu_5\omega_4^* - \frac23\nu_5\omega_5^*,\\
&16\pi^2\beta_{\kappa_2} =  \kappa_1\left(2\nu_2 + 2\omega_2 \right) \\
&\qquad\qquad\quad +\kappa_2\left(2\lambda_3 + 2\nu_1 + 2\nu_2 + 2\omega_1 + 2\omega_2 + \frac83\mu_1 + \frac83\mu_2 + \mu_3 + 8\mu_4 + \mu_5 - \frac13\mu_6\right)\\
&\qquad\qquad\quad + 2\kappa_2^*\lambda_5 + 4\kappa_3^*\nu_3 + 4\kappa_3\omega_3
+ \frac{17}3\nu_4\omega_4^* +\frac43\nu_4\omega_5^* +\frac43\nu_5\omega_4^* + \frac{17}3\nu_5\omega_5^* ,\\
&16\pi^2\beta_{\kappa_3} = \kappa_3\left(2\lambda_3 + 2\lambda_4 + 2\nu_1 + 3\nu_2 + 2\omega_1 + 3\omega_2 - \frac13\mu_1 - \frac13\mu_2 + \mu_3 + \mu_4 + 8\mu_5 + \frac83\mu_6\right)\\
&\qquad\qquad\quad + \nu_3\left(4\kappa_1^* + 6\kappa_2^*\right) + \omega_3\left(4\kappa_1 + 6\kappa_2\right)
+\frac{17}6\nu_4\omega_4 + \frac46\nu_4\omega_5 + \frac46\nu_5\omega_4 + \frac{17}6\nu_5\omega_5,\\
&16\pi^2\beta_{\mu_1} =  3\nu_4\nu_5^* + 3\omega_4\omega_5^* + 7\mu_1^2 + \mu_1\left(6\mu_2 + 6\mu_3 + 4\mu_4 - \mu_5 - 2\mu_6\right) \\
&\qquad\qquad\quad + \mu_2\left(4\mu_4 - \mu_5\right)  - 2\mu_4\mu_6 + 2\mu_5\mu_6 + \mu_6^2,\\
&16\pi^2\beta_{\mu_2} =  3\nu_4^*\nu_5 + 3\omega_4^*\omega_5 + 7\mu_2^2 + \mu_1\left(6\mu_2 + 4\mu_4 - \mu_5 \right) \\
&\qquad\qquad\quad + \mu_2\left(6\mu_3  + 4\mu_4 - \mu_5 - 2\mu_6\right) - 2\mu_4\mu_6 + 2\mu_5\mu_6 + \mu_6^2,\\
&16\pi^2\beta_{\mu_3} = 2\nu_1^2 + 2\nu_1\nu_2 + 2\omega_1^2 + 2\omega_1\omega_2 + 4\left|\kappa_1\right|^2 + 2\kappa_1\kappa_2^* + 2\kappa_1^*\kappa_2\\
&\qquad\qquad\quad - \frac13\left(\left|\nu_4\right|^2 + 4\nu_4\nu_5^* + 4\nu_4^*\nu_5 + \left|\nu_5\right|^2\right) - \frac13\left(\left|\omega_4\right|^2 + 4\omega_4\omega_5^* + 4\omega_4^*\omega_5 + \left|\omega_5\right|^2\right)\\
&\qquad\qquad\quad + \frac{31}9\mu_1^2  + \mu_1\left( \frac{32}9\mu_2 + 16\mu_3 + \frac{16}3\mu_4 + \mu_5 + \frac{29}9\mu_6\right) + \frac{31}9\mu_2^2 \\
&\qquad\qquad\quad + \mu_2\left(16\mu_3 + \frac{16}3\mu_4 + \mu_5 + \frac{29}9\mu_6\right)
+ 20\mu_3^2  + \mu_3\left(20\mu_4 + 6\mu_5 + 10\mu_6\right)  \\
&\qquad\qquad\quad + 3\mu_4^2 + \mu_4\left(2\mu_5 + \frac{22}3\mu_6\right) + \mu_5^2 + \mu_5\mu_6 + \frac{29}{18}\mu_6^2,\\
&16\pi^2\beta_{\mu_4} = \nu_2^2 + \omega_2^2 + 2\left|\kappa_2\right|^2 \\
&\qquad\qquad\quad + \frac13\left(2\left|\nu_4\right|^2 - \nu_4\nu_5^* - \nu_4^*\nu_5 + 2\left|\nu_5\right|^2\right) + \frac13\left(2\left|\omega_4\right|^2 - \omega_4\omega_5^* - \omega_4^*\omega_5 + 2\left|\omega_5\right|^2\right)\\
&\qquad\qquad\quad + \frac19\mu_1^2 + \mu_1\left(\frac{26}9\mu_2 + \frac{16}3\mu_4 + \mu_5 - \frac49\mu_6\right) + \frac19\mu_2^2 + \mu_2\left(\frac{16}3\mu_4  + \mu_5 - \frac49\mu_6 \right) \\
&\qquad\qquad\quad + 6\mu_3\mu_4 + 10\mu_4^2 + \mu_4\left(2\mu_5 + \frac43\mu_6\right) + 4\mu_5^2 + \mu_5\mu_6 - \frac29\mu_6^2,\\
&16\pi^2\beta_{\mu_5} = 4\nu_3^2 + 4\omega_3^2 + 2\left|\kappa_3\right|^2  \\
&\qquad\qquad\quad + \frac13\left(2\left|\nu_4\right|^2 - \nu_4\nu_5^* - \nu_4^*\nu_5 + 2\left|\nu_5\right|^2\right) + \frac13\left(2\left|\omega_4\right|^2 - \omega_4\omega_5^* - \omega_4^*\omega_5 + 2\left|\omega_5\right|^2\right)\\
&\qquad\qquad\quad + \frac19\mu_1^2 +  \mu_1\left( - \frac{10}9\mu_2 + \frac13\mu_5 - \frac49\mu_6\right) + \frac19\mu_2^2 + \mu_2\left(\frac13\mu_5 - \frac49\mu_6\right) + 6\mu_3\mu_5
 \\
&\qquad\qquad\quad + \mu_4\left(8\mu_5 + 2\mu_6\right) + 8\mu_5^2  + \frac{19}3\mu_5\mu_6  + \frac79\mu_6^2,\\
&16\pi^2\beta_{\mu_6} =  3\nu_4\nu_5^* + 3\nu_4^*\nu_5 + 3\omega_4\omega_5^* + 3\omega_4^*\omega_5 \\
&\qquad\qquad\quad - 2\mu_1^2 + \mu_1\left(6\mu_5 + 7\mu_6 \right) - 2\mu_2^2 + \mu_2\left(6\mu_5 + 7\mu_6\right) + 6\mu_3\mu_6 + \frac12\mu_6^2.
\end{align*}
\endgroup


\clearpage

\begin{thebibliography}{999}

\bibitem{Manohar:2006ga} 
  A.~V.~Manohar and M.~B.~Wise,
  Phys.\ Rev.\ D {\bf 74}, 035009 (2006)
  [hep-ph/0606172].

\bibitem{Perez:2016qbo} 
  P.~Fileviez Perez and C.~Murgui,
  arXiv:1604.03377 [hep-ph].

\bibitem{Dorsner:2007fy} 
  I.~Dorsner and I.~Mocioiu,
  Nucl.\ Phys.\ B {\bf 796}, 123 (2008)
  doi:10.1016/j.nuclphysb.2007.12.004
  [arXiv:0708.3332 [hep-ph]].


\bibitem{Bertolini:2013vta} 
  S.~Bertolini, L.~Di Luzio and M.~Malinsky,
  Phys.\ Rev.\ D {\bf 87}, no. 8, 085020 (2013)
  doi:10.1103/PhysRevD.87.085020
  [arXiv:1302.3401 [hep-ph]].


\bibitem{Gresham:2007ri} 
  M.~I.~Gresham and M.~B.~Wise,
  Phys.\ Rev.\ D {\bf 76}, 075003 (2007)
  [arXiv:0706.0909 [hep-ph]].
  
\bibitem{Gerbush:2007fe} 
  M.~Gerbush, T.~J.~Khoo, D.~J.~Phalen, A.~Pierce and D.~Tucker-Smith,
  Phys.\ Rev.\ D {\bf 77}, 095003 (2008)
  [arXiv:0710.3133 [hep-ph]].



\bibitem{Burgess:2009wm} 
  C.~P.~Burgess, M.~Trott and S.~Zuberi,
  JHEP {\bf 0909}, 082 (2009)
  [arXiv:0907.2696 [hep-ph]].


\bibitem{He:2011ti} 
  X.~-G.~He and G.~Valencia,
  Phys.\ Lett.\ B {\bf 707}, 381 (2012)
  [arXiv:1108.0222 [hep-ph]].

\bibitem{Dobrescu:2011aa} 
  B.~A.~Dobrescu, G.~D.~Kribs and A.~Martin,
  Phys.\ Rev.\ D {\bf 85}, 074031 (2012)
  [arXiv:1112.2208 [hep-ph]].
  
\bibitem{Bai:2011aa} 
  Y.~Bai, J.~Fan and J.~L.~Hewett,
  JHEP {\bf 1208}, 014 (2012)
  [arXiv:1112.1964 [hep-ph]].
 
\bibitem{Arnold:2011ra} 
  J.~M.~Arnold and B.~Fornal,
  Phys.\ Rev.\ D {\bf 85}, 055020 (2012)
  [arXiv:1112.0003 [hep-ph]].
 
  
\bibitem{Kribs:2012kz} 
  G.~D.~Kribs and A.~Martin,
  arXiv:1207.4496 [hep-ph].
  
\bibitem{Reece:2012gi} 
  M.~Reece,
  arXiv:1208.1765 [hep-ph].
  
\bibitem{Cao:2013wqa} 
  J.~Cao, P.~Wan, J.~M.~Yang and J.~Zhu,
  arXiv:1303.2426 [hep-ph].

\bibitem{He:2013tla} 
  X.~G.~He, H.~Phoon, Y.~Tang and G.~Valencia,
  JHEP {\bf 1305}, 026 (2013)
  doi:10.1007/JHEP05(2013)026
  [arXiv:1303.4848 [hep-ph]].


\bibitem{Cheng:2015lsa} 
  X.~D.~Cheng, X.~Q.~Li, Y.~D.~Yang and X.~Zhang,
  J.\ Phys.\ G {\bf 42}, no. 12, 125005 (2015)
  doi:10.1088/0954-3899/42/12/125005
  [arXiv:1504.00839 [hep-ph]].


\bibitem{Martinez:2016fyd} 
  R.~Martinez and G.~Valencia,
  arXiv:1612.00561 [hep-ph].



\bibitem{Cheng:2016tlc} 
  L.~Cheng and G.~Valencia,
  JHEP {\bf 1609}, 079 (2016)
  doi:10.1007/JHEP09(2016)079
  [arXiv:1606.01298 [hep-ph]].



\bibitem{Chakrabarty:2014aya} 
  N.~Chakrabarty, U.~K.~Dey and B.~Mukhopadhyaya,
  JHEP {\bf 1412}, 166 (2014)
  doi:10.1007/JHEP12(2014)166
  [arXiv:1407.2145 [hep-ph]].

\bibitem{Ferreira:2015rha} 
  P.~Ferreira, H.~E.~Haber and E.~Santos,
  Phys.\ Rev.\ D {\bf 92}, 033003 (2015)
  Erratum: [Phys.\ Rev.\ D {\bf 94}, no. 5, 059903 (2016)]
  doi:10.1103/PhysRevD.92.033003, 10.1103/PhysRevD.94.059903
  [arXiv:1505.04001 [hep-ph]].
  
\bibitem{Callaway:1988ya} 
  D.~J.~E.~Callaway,
  Phys.\ Rept.\  {\bf 167}, 241 (1988).
  doi:10.1016/0370-1573(88)90008-7
  
\bibitem{Sher:1988mj} 
  M.~Sher,
  Phys.\ Rept.\  {\bf 179}, 273 (1989).
  doi:10.1016/0370-1573(89)90061-6

\bibitem{Gunion:1989we} 
  J.~F.~Gunion, H.~E.~Haber, G.~L.~Kane and S.~Dawson,
  Front.\ Phys.\  {\bf 80}, 1 (2000).
 
\bibitem{Branco:2011iw} 
  G.~C.~Branco, P.~M.~Ferreira, L.~Lavoura, M.~N.~Rebelo, M.~Sher and J.~P.~Silva,
  Phys.\ Rept.\  {\bf 516}, 1 (2012)
  [arXiv:1106.0034 [hep-ph]].

\bibitem{Chivukula:1987py}
  R.~S.~Chivukula and H.~Georgi,
  Phys.\ Lett.\ B {\bf 188} (1987) 99.
  
\bibitem{D'Ambrosio:2002ex}
  G.~D'Ambrosio, G.~F.~Giudice, G.~Isidori and A.~Strumia,
  Nucl.\ Phys.\ B {\bf 645} (2002) 155
  [hep-ph/0207036].


\bibitem{Sikivie:1980hm} 
  P.~Sikivie, L.~Susskind, M.~B.~Voloshin and V.~I.~Zakharov,
  Nucl.\ Phys.\ B {\bf 173}, 189 (1980).
  doi:10.1016/0550-3213(80)90214-X
  
\bibitem{Pomarol:1993mu} 
  A.~Pomarol and R.~Vega,
  Nucl.\ Phys.\ B {\bf 413}, 3 (1994)
  doi:10.1016/0550-3213(94)90611-4
  [hep-ph/9305272].

\bibitem{Grzadkowski:2010dj} 
  B.~Grzadkowski, M.~Maniatis and J.~Wudka,
  JHEP {\bf 1111}, 030 (2011)
  doi:10.1007/JHEP11(2011)030
  [arXiv:1011.5228 [hep-ph]].

\bibitem{Gerard:2007kn} 
  J.-M.~Gerard and M.~Herquet,
  Phys.\ Rev.\ Lett.\  {\bf 98}, 251802 (2007)
  doi:10.1103/PhysRevLett.98.251802
  [hep-ph/0703051 [HEP-PH]].
 
\bibitem{Cervero:2012cx} 
  E.~Cervero and J.~M.~Gerard,
  Phys.\ Lett.\ B {\bf 712}, 255 (2012)
  doi:10.1016/j.physletb.2012.05.010
  [arXiv:1202.1973 [hep-ph]].



\bibitem{Lee:1977eg} 
  B.~W.~Lee, C.~Quigg and H.~B.~Thacker,
  Phys.\ Rev.\ D {\bf 16}, 1519 (1977).
  
\bibitem{Kanemura:1993hm} 
  S.~Kanemura, T.~Kubota and E.~Takasugi,
  Phys.\ Lett.\ B {\bf 313}, 155 (1993)
  [hep-ph/9303263].

\bibitem{Horejsi:2005da} 
  J.~Horejsi and M.~Kladiva
  Eur.\ Phys.\ J.\ C {\bf 46}, 81 (2006)
  doi:10.1140/epjc/s2006-02472-3
  [hep-ph/0510154].

\bibitem{Ginzburg:2005dt} 
  I.~F.~Ginzburg and I.~P.~Ivanov,
  Phys.\ Rev.\ D {\bf 72}, 115010 (2005)
  doi:10.1103/PhysRevD.72.115010
  [hep-ph/0508020].
  
 
\bibitem{Grinstein:2015rtl} 
  B.~Grinstein, C.~W.~Murphy and P.~Uttayarat,
  arXiv:1512.04567 [hep-ph].


\bibitem{Deshpande:1977rw} 
  N.~G.~Deshpande and E.~Ma,
  Phys.\ Rev.\ D {\bf 18}, 2574 (1978).
  doi:10.1103/PhysRevD.18.2574
  
\bibitem{Kannike:2016fmd} 
  K.~Kannike,
  Eur.\ Phys.\ J.\ C {\bf 76}, no. 6, 324 (2016)
  doi:10.1140/epjc/s10052-016-4160-3
  [arXiv:1603.02680 [hep-ph]].

\bibitem{Hayreter:2017wra} 
  A.~Hayreter and G.~Valencia,
  arXiv:1703.04164 [hep-ph].

\bibitem{Marciano:1989ns} 
  W.~J.~Marciano, G.~Valencia and S.~Willenbrock,
  Phys.\ Rev.\ D {\bf 40}, 1725 (1989).
  
\bibitem{Cacchio:2016qyh} 
  V.~Cacchio, D.~Chowdhury, O.~Eberhardt and C.~W.~Murphy,
  JHEP {\bf 1611}, 026 (2016)
  doi:10.1007/JHEP11(2016)026
  [arXiv:1609.01290 [hep-ph]].
  
  
\bibitem{Murphy:2017ojk} 
  C.~W.~Murphy,
  arXiv:1702.08511 [hep-ph].
  

\bibitem{method}
Cook, D. and Buja, A. and Cabrera, J. and Hurley, C.,
Journal of Computational and Graphical Statistics 3, 155 (1995).
 
\bibitem{visual}
 Deborah F. Swayne and Duncan Temple Lang and Andreas Buja and Dianne Cook,
The method reference:



  
\end{thebibliography}
\end{document}